\documentclass{aa}
\usepackage[varg]{txfonts}
\usepackage{hyperref}
\usepackage{amsmath}
\usepackage{relsize}
\usepackage{dsfont}
\begin{document}
\title{The Spatial Evolution of Young Massive Clusters} 
\subtitle{I. A New Tool to Quantitatively Trace Stellar Clustering}

\author{Anne S.M. Buckner\inst{1}, Zeinab Khorrami\inst{2}, Pouria Khalaj\inst{3}, Stuart L. Lumsden\inst{1}, Isabelle Joncour\inst{3,5}, Estelle Moraux\inst{3}, Paul Clark \inst{2}, Ren\'e D. Oudmaijer\inst{1}, Jos\'e Manuel Blanco\inst{4}, Ignacio de la Calle\inst{4}, Jos\'e M. Herrera-Fernandez\inst{4}, Fr\'ed\'erique Motte\inst{3}, Jes\'us J. Salgado\inst{4} and Luis Valero-Mart\'in\inst{4}}

\institute{School of Physics and Astronomy, University of Leeds, Leeds LS2 9JT, U.K. \\ \email{a.s.m.buckner@leeds.ac.uk}
\and School of Physics and Astronomy, Cardiff University, The Parade, CF24 3AA, U.K. \and Universit\'e Grenoble Alpes, CNRS, IPAG, 38000 Grenoble, France \and Quasar Science Resources, S.L., Edificio Ceudas, Ctra. de La Coru\~na, Km 22.300, 28232, Las Rozas de Madrid, Madrid, Spain \and Department of Astronomy, University of Maryland, College Park, MD 20742, USA}

\date{Received 2 March 2018 / Accepted 8 January 2019}

\abstract{There are a number of methods that identify stellar sub-structure in star forming regions, but these do not quantify the degree of association of individual stars - something which is required if we are to better understand the mechanisms and physical processes that dictate structure.}{To present the new novel statistical clustering tool “INDICATE” which assesses and quantifies the degree of spatial clustering of each object in a dataset, discuss its applications as a tracer of morphological stellar features in star forming regions, and to look for these features in the Carina Nebula (NGC\,3372).}  {We employ a nearest neighbour approach to quantitatively compare the spatial distribution in the local neighbourhood of an object with that expected in an evenly spaced uniform (i.e. definitively non-clustered) field. Each object is assigned a clustering index (‘$I$’) value, which is a quantitative measure of its clustering tendency. We have calibrated our tool against random distributions to aid interpretation and identification of significant $I$ values.}{Using INDICATE we successfully recover known stellar structure of the Carina Nebula, including the young Trumpler 14-16, Treasure Chest and Bochum\,11 clusters. Four sub-clusters contain no, or very few, stars with a degree of association above random which suggests these sub-clusters may be fluctuations in the field rather than real clusters. In addition we find: (1) Stars in the NW and SE regions have significantly different clustering tendencies, which is reflective of differences in the apparent star formation activity in these regions. Further study is required to ascertain the physical origin of the difference; (2) The different clustering properties between the NW and SE regions are also seen for OB stars and are even more pronounced; (3)  There are no signatures of classical mass segregation present in the SE region - massive stars here are not spatially concentrated together above random;  (4)  Stellar concentrations are more frequent around massive stars than typical for the general population, particularly in the Tr14 cluster; (5) There is a relation between the concentration of OB stars and the concentration of (lower mass) stars around OB stars in the centrally concentrated Tr14 and Tr15, but no such relation exists in Tr16. We conclude this is due to the highly sub-structured nature of Tr16.}{INDICATE is a powerful new tool employing a novel approach to quantify the clustering tendencies of individual objects in a dataset within a user-defined parameter space. As such it can be used in a wide array of data analysis applications. In this paper we have discussed and demonstrated its application to trace morphological features of young massive clusters.}


\titlerunning{The Spatial Evolution of Young Massive Clusters I}
\authorrunning{Buckner et al.}

\maketitle

%######################################################################################################

\section{Introduction}

Massive stars are fundamental to the evolution of galaxies, profoundly
impacting the interstellar medium through chemical enrichment (outflows,
supernovae), mixing and turbulence (winds, outflows, supernovae), and
heating/cooling (ionising radiation).  

Unfortunately while isolated low mass star formation appears to be
  well described observationally (e.g. \citealt{1987ARA&A..25...23S}, \citealt{2000prpl.conf...59A}, 
  \citealt{2012ARA&A..50...65L}), there is still little consensus about the
  formation of massive stars.  This is largely due to observational reasons.
  High mass stars are rare, evolve rapidly, and have shorter lifetimes than
  low mass stars.  They also emerge onto the main sequence still heavily
  embedded, having formed almost exclusively in associations, groups and
  clusters \citep{2005A&A...437..247D}.  The linked formation and evolution of both massive stars and
  clusters, and how they interact is clearly part of the ``picture'' for
  massive star formation, but much is still unknown (e.g. \citealt{2007ARA&A..45..481Z}).

To discriminate between different models for cluster and/or massive
  star formation/evolution requires a multi-pronged analysis of the structure
  and dynamics of the stars and gas in Young Massive
Clusters (YMCs). To this end we created the
StarFormMapper\footnote{\url{https://starformmapper.org/}} (SFM) project.  

One of the fundamental analytical techniques required is to study how the stars
and gas ``cluster'' together.  Here, we are particularly interested in the
study of the intensity, correlation and spatial distribution of point
processes, which collectively help to define the distribution and clustering of
those points (see, e.g. \citealt{MollerWaagepetersen2007ScanJStats}).  We are not
concerned in this paper with searching for stellar ``sub-structure'' (discrete star groupings), but rather for
suitable statistical measures of the distribution of these point patterns.
This is complicated in star formation regions as the
distributions of stars and gas are inherently heterogeneous.  Many of the best
understood statistics from other fields are therefore not easily applied
(or are simply invalid).  In addition, we wish to use techniques which are
valid in any number of dimensions and are applicable easily at different
distances, whilst still being computationally simple.

Several global methods have been used in the past.  The 2-point correlation
function is well studied in cosmology, but has also been used in star forming
regions (e.g. \citealt{gomez_spatial_1993}, \citealt{ScaloChappell1999}).
The Q parameter by \citet{2004MNRAS.348..589C} uses a very different technique,
which compares the average length from the Minimum Spanning Tree (MST) with the
average length from the complete graph of all points, and can distinguish
between a smooth overall radial density gradient and multi-scale fractal
sub-clustering in a region.  It has
successfully identified signatures of sub-structure in the Cygnus OB2
\citep{2014MNRAS.438..639W}, Serpens, Ophiucius and Perseus star forming
regions \citep{schmeja_structures_2008} and has been applied to assess the
dynamical status of star clusters in numerical simulations
(e.g. \citealt{parker_dynamical_2014}).  Similar methods have also been applied
to the study of mass segregation (see e.g. \citealt{parker_comparisons_2015}).  However these methods still suffer if heterogeneous
structures are present (e.g. \citealt{2009A&A...503..909C}).  In particular if we wish
to compare observations and simulations, great care must be taken in such
circumstances. 

An alternative possible approach is suggested by the field of geostatistics
where interest has also focused on the use of {\em local indicators} (e.g. 
\citealt{Anselin1995LISApaper}).  In this case, rather than calculating a single
parameter for a group of stars/gas as a whole, every unique point has its own
derived value.  These can then be used to characterise the distribution. That
is the approach we will follow here.

In this paper we present our new statistical clustering tool “INDICATE” (INdex
to Define Inherent Clustering And TEndencies) that we are currently
implementing in the SFM project.  \citet{10.2307/42907238} established the
Hopkins statistic to assess the global clustering tendency of a dataset by
testing its spatial randomness through quantitative measurements of its
uniformity. A single global value is calculated through a comparison of the mean
k-nearest neighbour distances between objects within the dataset, and between
points in the dataset and a similarly constructed uniform random sample.  We
propose instead to derive a similar index but for {\em every} point in
the dataset individually.  
 
This paper is structured as follows. Section \ref{sect_method} describes how
our tool works and is calibrated. In Section \ref{sec_obs} we use INDICATE to
trace stellar morphological features of star formation, demonstrating its
ability to cope with the complex, often poorly defined, spatial clusterings
expected in young massive star forming regions/clusters. Our conclusions are presented in Section \ref{sec_conclude}.

%######################################################################################################

\begin{figure*}
\centering
   \includegraphics[width=0.45\textwidth]{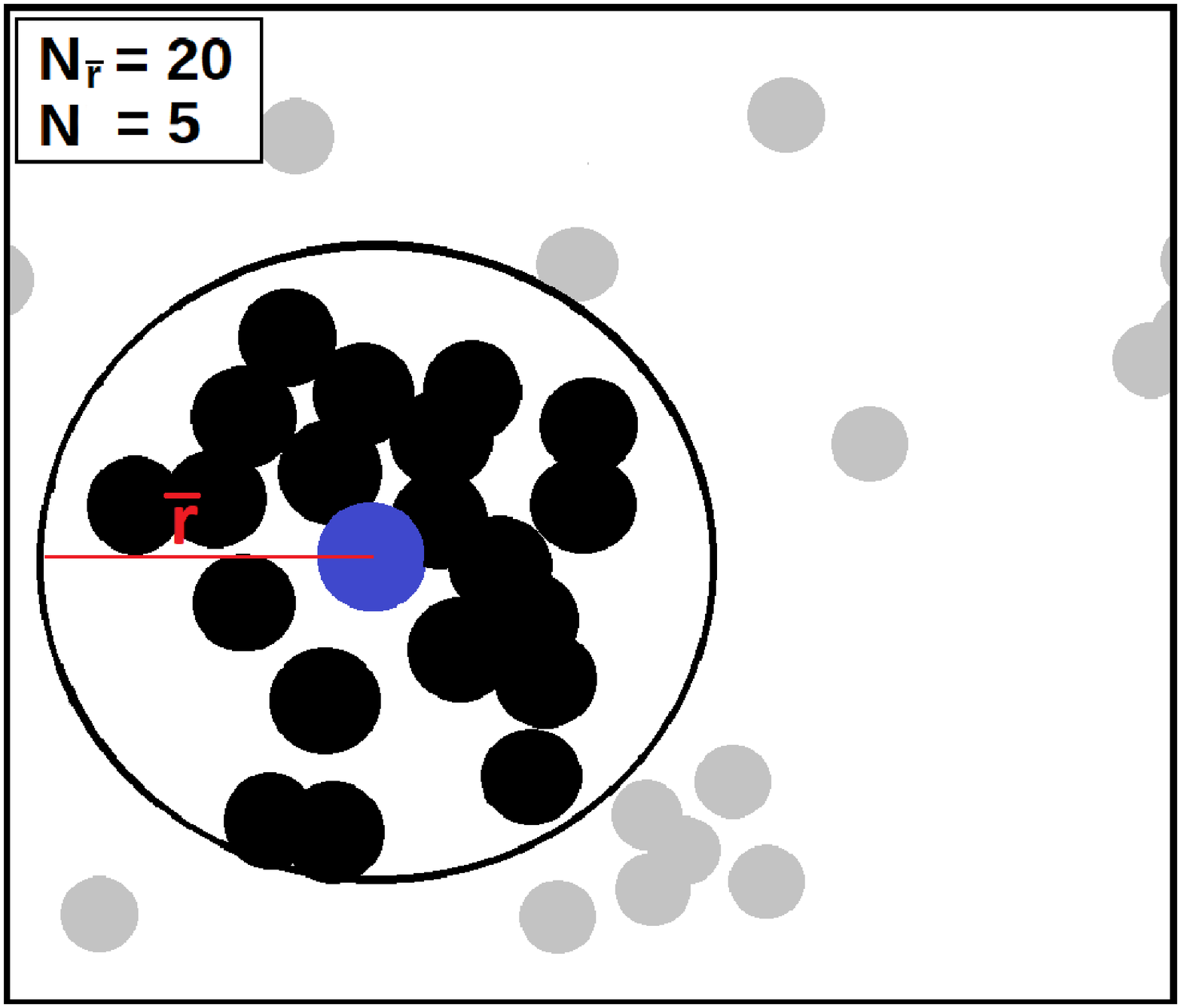} \hfill
   \includegraphics[width=0.45\textwidth]{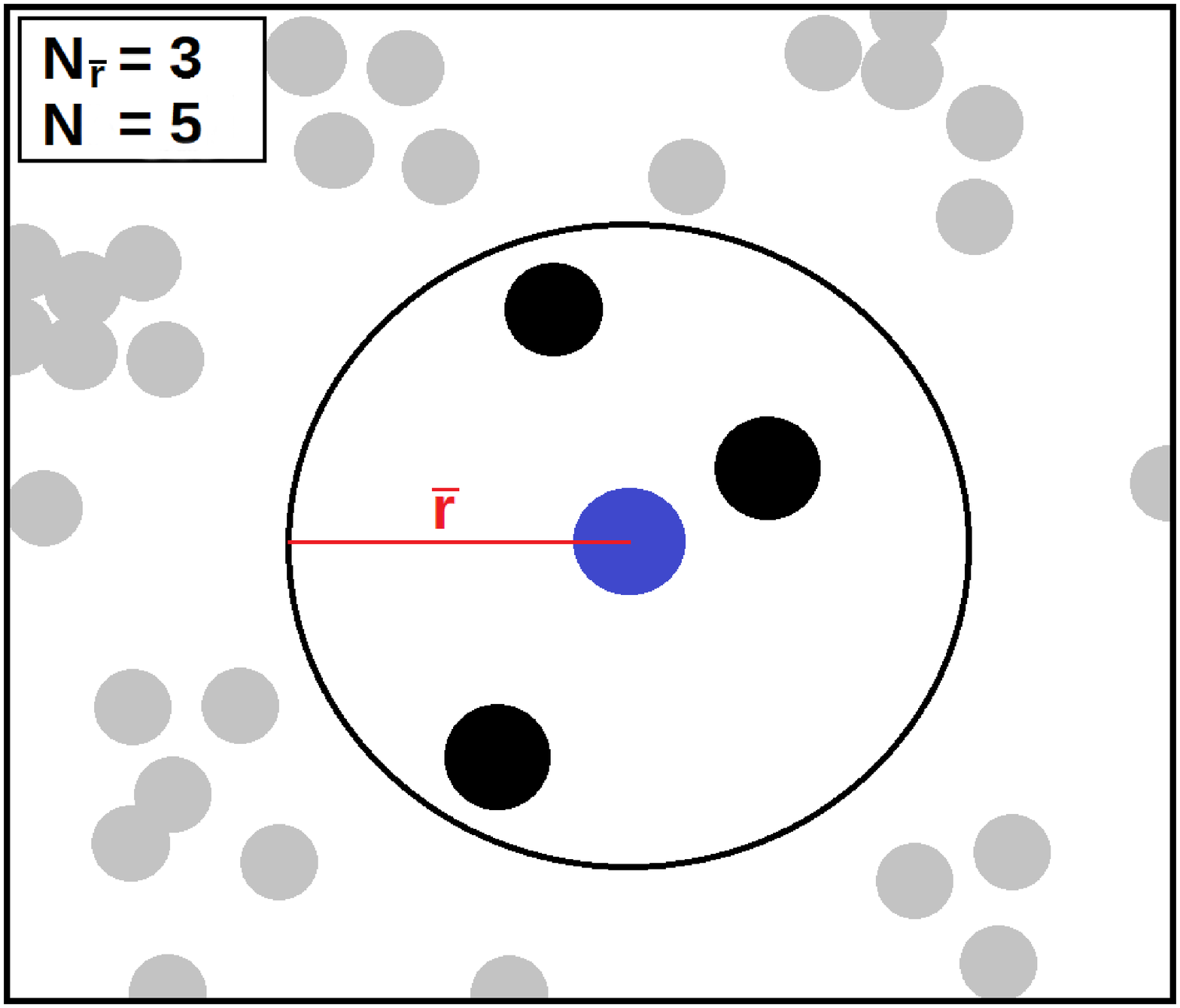} \hfill

  \caption{Demonstration of how INDICATE defines the index $I_{j,\mathit{ N}}$ for a point. All points within a radius of $\bar{r}$ of the selected point (marked in blue) are counted ($N_\mathrm{\bar{r}}$) and compared to the number of points expected within the same radius in an evenly spaced uniform point distribution with the same number density as the points parent sample ($\mathit{N}$). The index of the blue point is calculated using Eq.\,\ref{eq_I} \,as (Left:) $I_5=4.0$ and (Right:) $I_5=0.6$.}  \label{fig_howworks}
\end{figure*}

\section{INDICATE: INdex to Define Inherent Clustering And TEndencies}\label{sect_method}

\subsection{General Tool Description}\label{sect_method_descrip}

INDICATE is a tool to quantify the degree of association of each point in a 2+D
discrete dataset.  It requires no a priori knowledge of -- nor makes
assumptions about -- the sub-structure present in a dataset, since it is a {\em
  local} statistic.  The separation of the spatial position of the $j^{th}$
point in the actual dataset with the $N^{th}$ nearest neighbour in an evenly
spaced uniform (i.e. definitively non-clustered) control distribution is
determined. The mean value of this separation, $\bar{r}$, is then derived.  Finally, INDICATE assigns an index, $I$, to
every point in the actual dataset, which is simply the ratio of the actual number of real neighbours
within this mean separation, $\bar{r}$, and $N$, the nearest neighbour
number.  Since our tool fundamentally relies on properties linked to distance ratios
in the real and uniform sample, it is itself region distance independent in principle
at least (but see Appendix \ref{sec_inter}).  Because our tool uses an evenly spaced grid as the
comparison it is computationally less intensive than a direct implementation of
the Hopkins, or similar, statistics.  Below we describe step-by-step how this
index is derived for a simple 2D distribution.  A future paper will deal with
the implementation to a full 3D dataset using Gaia parallaxes.

%~~~~~~~~~~~~~~~~~~~~~~~~~~~~~~~~~~~~~~~~~~~~~~~~~~~~~~~~~~~~~~~~~~~~
\subsubsection{Step I: Define the bounds of the dataset}

\textit{INDICATE is designed to be applicable to any desired
    N-dimensional parameter space.  However, as outlined here, it assumes that
    all dimensions have the same scaling (e.g. J2000 sky coordinates should be
    converted to a local coordinate frame prior to beginning).}

The bounds of the dataset parameter space are defined from the density
  distribution, and the area occupied by the data, $A$, measured. The shape of
  the delimited area has a negligible effect on the tool (Appendix\,\ref{sec_edge}), but for clarity we will use a
  rectangular parameter space for all our explanatory datasets in the
  description of the tool. In practice, the shape is related to the problem to be studied, and is defined by the user. 

The number density, $n_{obs}$, of the dataset is determined using,

\begin{equation}
\\ n_\mathrm{obs}=\frac{N_\mathrm{tot}}{A}
\end{equation}

where $N_\mathrm{tot}$ is the total number of points in the dataset.

%~~~~~~~~~~~~~~~~~~~~~~~~~~~~~~~~~~~~~~~~~~~~~~~~~~~~~~~~~~~~~~~~~~~~
\subsubsection{Step II: Generate the control distribution}

An adaptive evenly spaced uniform point distribution which we designate the \textit{control distribution} is generated. The control distribution is rectangular (regardless of dataset shape defined in Step I), populates the bounded parameter space and has the same number density as the dataset i.e.

		\begin{equation}\label{eq_ncon}
		\\	n_\mathrm{con} = n_\mathrm{obs}
		\end{equation}

		%~~~~~~~~~~~~~~~~~~~~~~~~~~~~~~~~~~~~~~~~~~~~~~~~~~~~~~~~~~~~~~~~~~~~
\subsubsection{Step III: Measure the mean Nearest Neighbour distance}\label{nn_dist}

	The mean Euclidean distance, $\bar{r}$, of each point, $j$, in the dataset to its N$^{th}$ nearest neighbour in the control distribution is measured using:
	
	\begin{equation}\label{eq_rbar}
	\\ \bar{r}= \frac{\sum\limits_{j=1}^{N_\mathrm{tot}}\,r_{j}}{N_\mathrm{tot}}
	\end{equation}

where,
	\begin{equation}\label{eq_rj}
	\\ r_{j} = \sqrt{(x_j-x^\mathrm{con}_\mathit{N})^2 +(y_j-y^\mathrm{con}_\mathit{N})^2} 
	\end{equation}

and ($x_j$, $y_j$) are the respective x and y axis coordinates of point $j$ and ($x^\mathrm{con}_\mathit{N}$, $y^\mathrm{con}_\mathit{N}$)  are the respective x and y axis coordinates of the N$^{th}$ nearest neighbour in the control distribution.

		%~~~~~~~~~~~~~~~~~~~~~~~~~~~~~~~~~~~~~~~~~~~~~~~~~~~~~~~~~~~~~~~~~~~~
\subsubsection{Step IV: Calculate the Index, $I$}\label{sect_calc_I} 

The number of points, $N_\mathrm{\bar{r}}$, closer than $\bar{r}$ to each point, $j$, in the dataset is counted (see Figure\,\ref{fig_howworks}). The index of point $j$ is then the ratio of the number of neighbours closer than $\bar{r}$ in the dataset with that expected by a non-clustered distribution i.e.

			\begin{equation}\label{eq_I}
			\\ I_{j,\mathit{ N}} = \frac{N_\mathrm{\bar{r}}}{\mathit{N}}
			\end{equation}

where $\mathit{N}$ is the N$^{th}$ nearest neighbour number (e.g. if $\bar{r}$ is measured for the 5$^{th}$ nearest neighbour, $\mathit{N}=5$; 6$^{th}$ nearest neighbour, $\mathit{N}=6$...etc.). The ratio $I_{j,\mathit{ N}}$ is unitless and has a range of $0 \leq I_{j,\mathit{ N}} \leq \frac{N_\mathrm{tot}- 1}{\mathit{N}}$, such that the higher its value the more spatially clustered point $j$ is. 

\begin{figure*}
\centering
   \includegraphics[width=6cm,height=5cm]{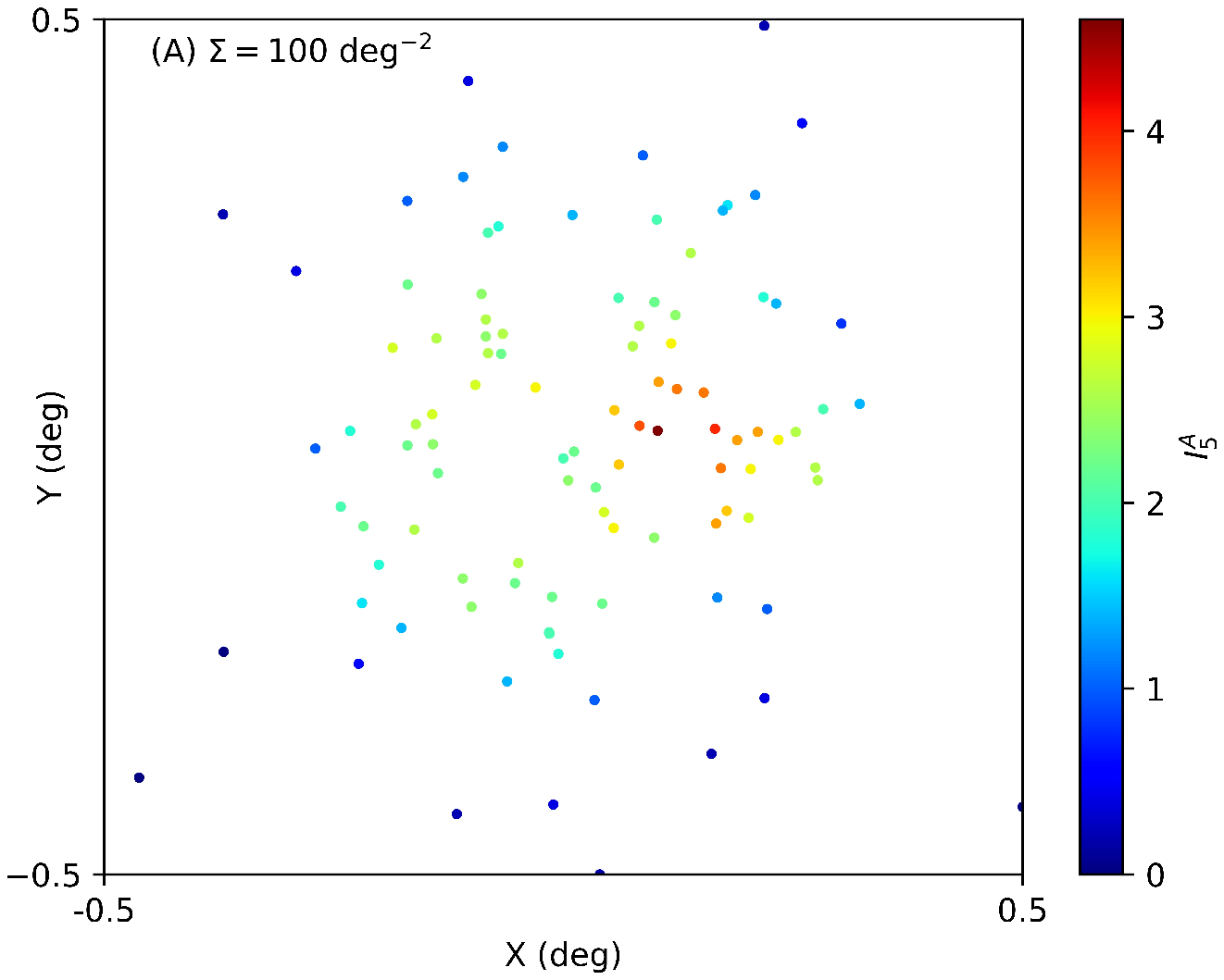} \hfill
   \includegraphics[width=6cm,height=5cm]{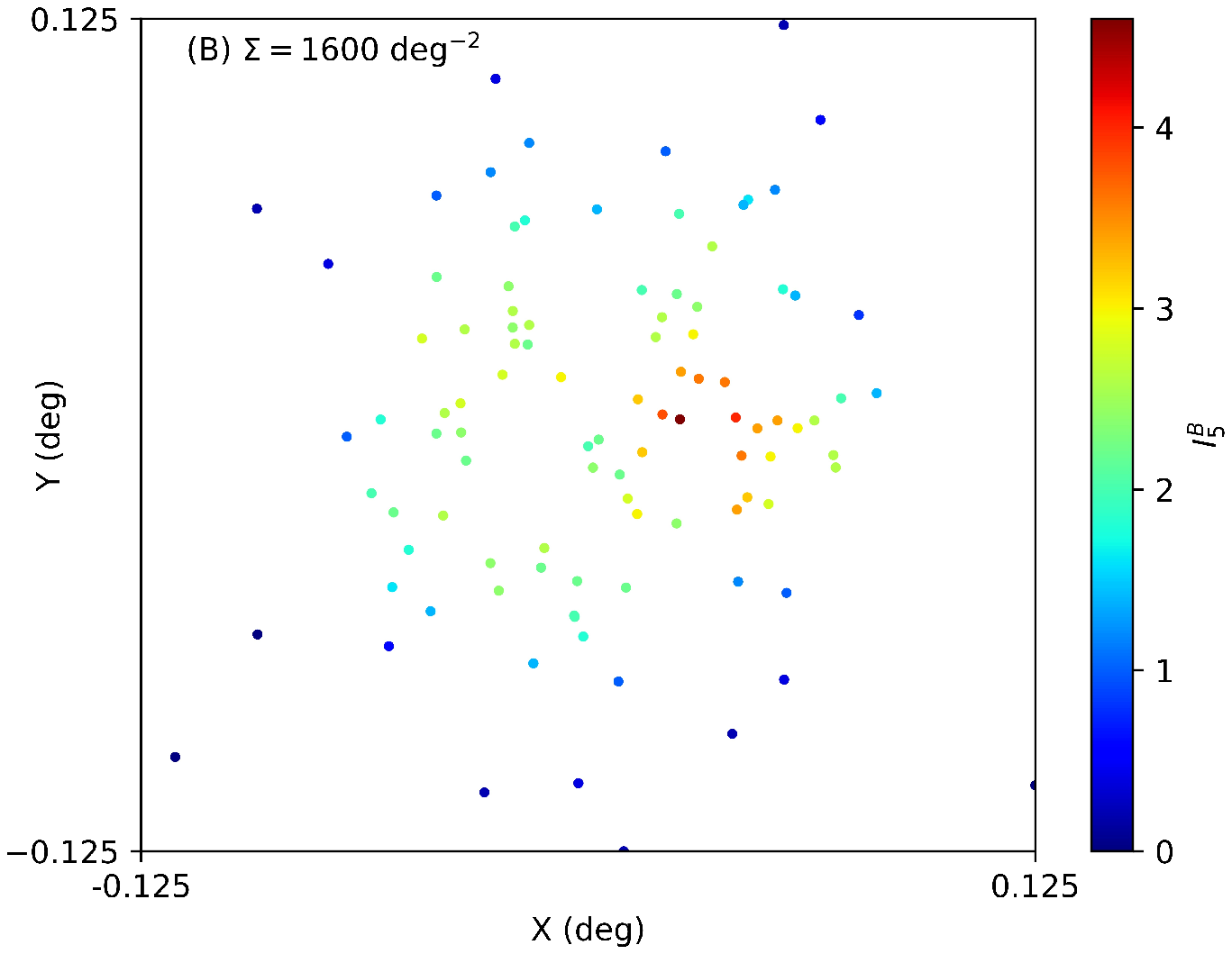} \hfill
   \includegraphics[width=6cm,height=5cm]{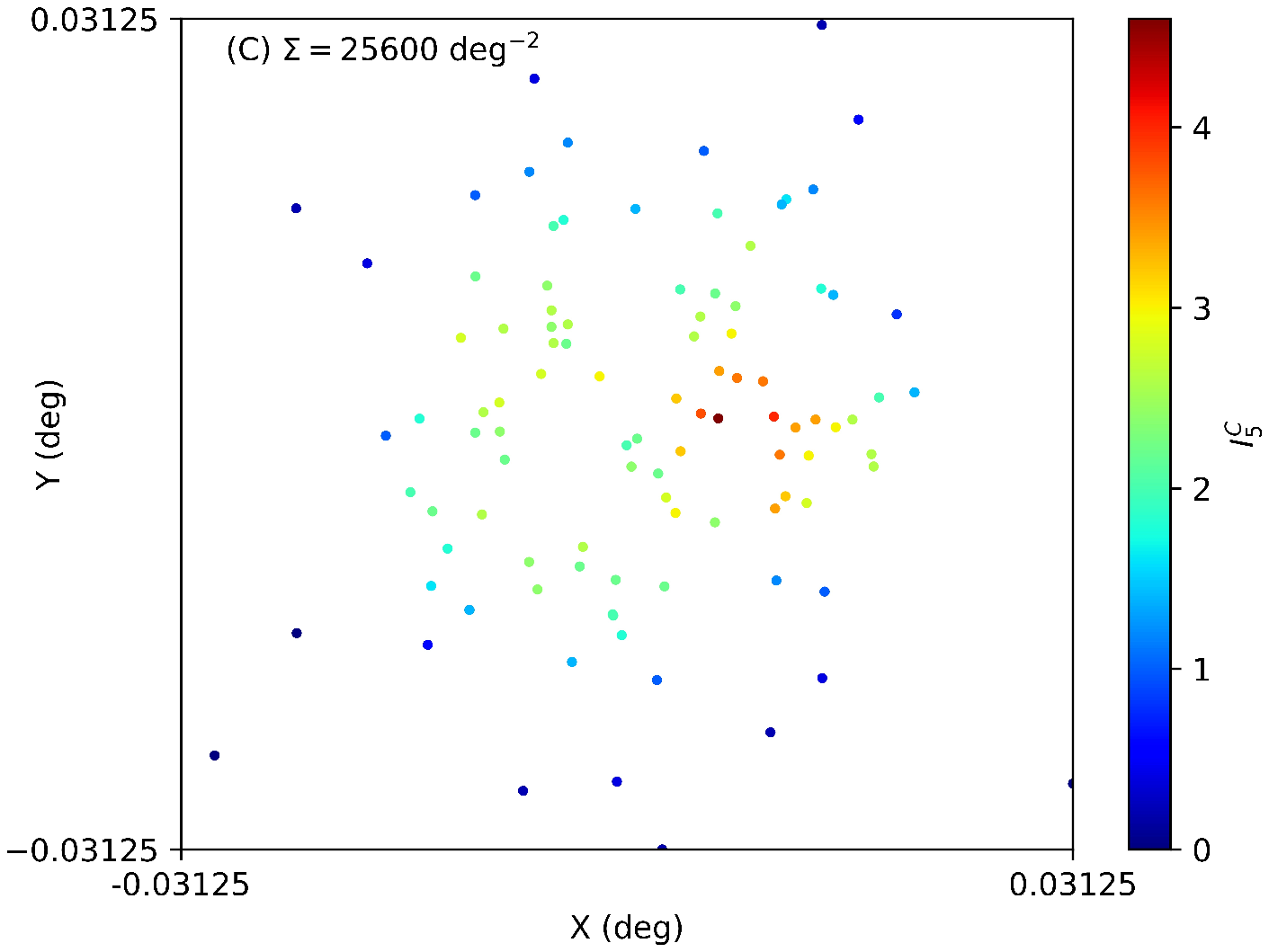} \hfill
  \caption{Plots A-C show the index values derived by INDICATE for a synthetic Gaussian cluster with 100 members, using a nearest neighbour number of $\mathit{N}=5$ in all three instances. In B and C the angular dispersion of the cluster was reduced to simulate how it would be observed if its distance was a factor of 4 and 16 times larger than A, respectively. Despite the significant increase in point surface densities, the index values derived for each star is unchanged because there is no change in the relative spatial distribution of members.}  \label{fig_distance} 
\end{figure*}

It is important to note the index is not a measure of local surface density - Eq.\,\ref{eq_I} describes the \textit{local spatial distribution} of point $j$. Therefore although the index is proportional to the local point surface density of a dataset, it is possible for two datasets with significantly different densities to have identical index values if their points have the same spatial distribution. For example, Figure\,\ref{fig_distance}A shows index values derived for Gaussian cluster with 100 members, using a nearest neighbour number of $\mathit{N}=5$. On visual inspection the highest values have been assigned to stars with the highest degree of association. Figures \,\ref{fig_distance}B and \,\ref{fig_distance}C show the same cluster, but with an observed angular dispersion and surface density the cluster would have if its distance was a factor of 4 and 16 times larger than A respectively. Applying INDICATE under the same conditions as for \,\ref{fig_distance}A, the index values for each star remain unchanged in \ref{fig_distance}B and \,\ref{fig_distance}C i.e. for all members $\Delta\,I_{5} \equiv 0$ despite the surface density increasing by a factor of 256 between \ref{fig_distance}A and \ref{fig_distance}C, as the local spatial distribution of members in all three clusters is identical. 

Hence the index can be used to directly analyse variation in the spatial distribution of points, in any desired parameter space, (a) within a dataset and/or (b) comparatively between two or more datasets.  

\begin{figure*}
\centering
   \includegraphics[width=6cm,height=5cm]{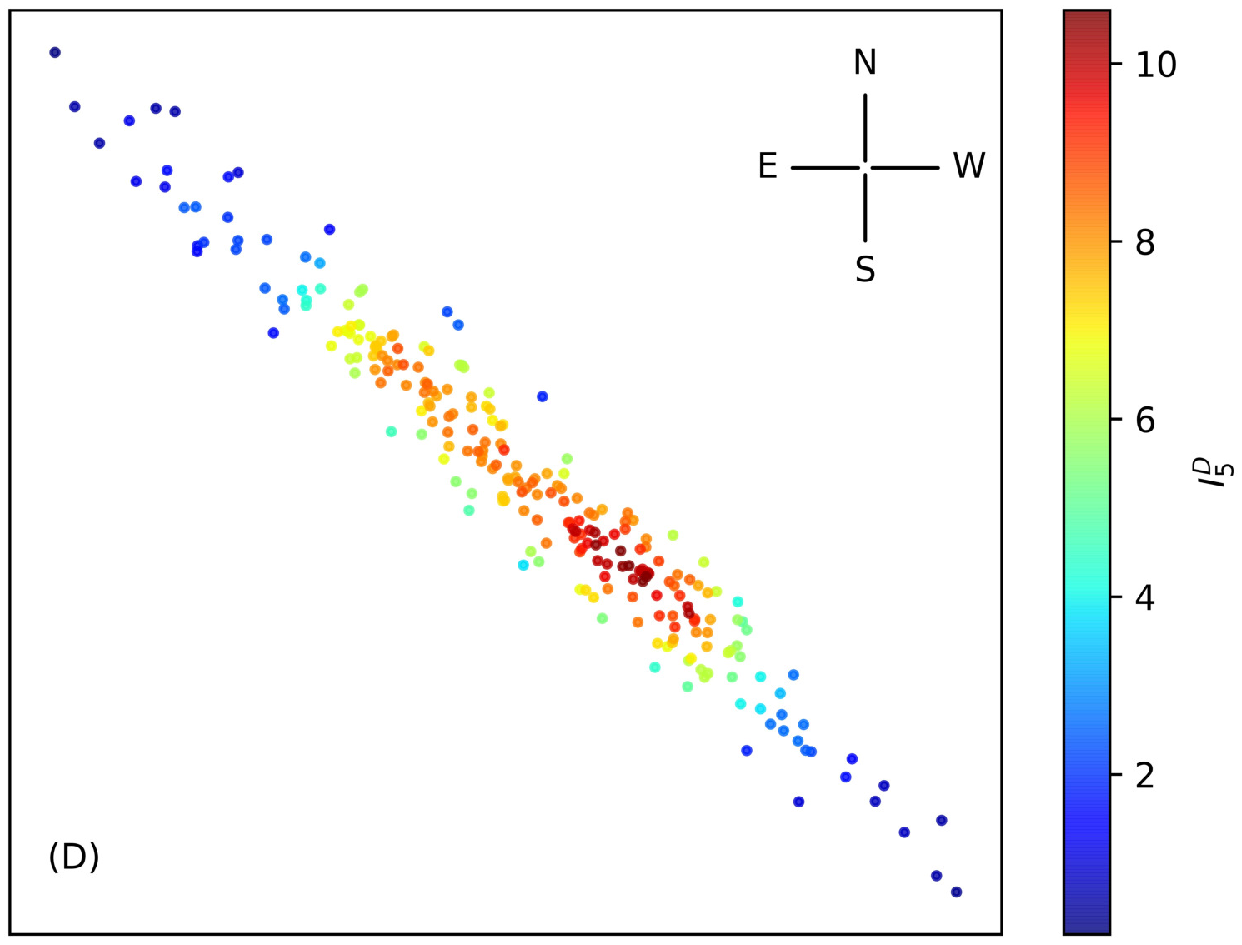} \hfill
   \includegraphics[width=6cm,height=5cm]{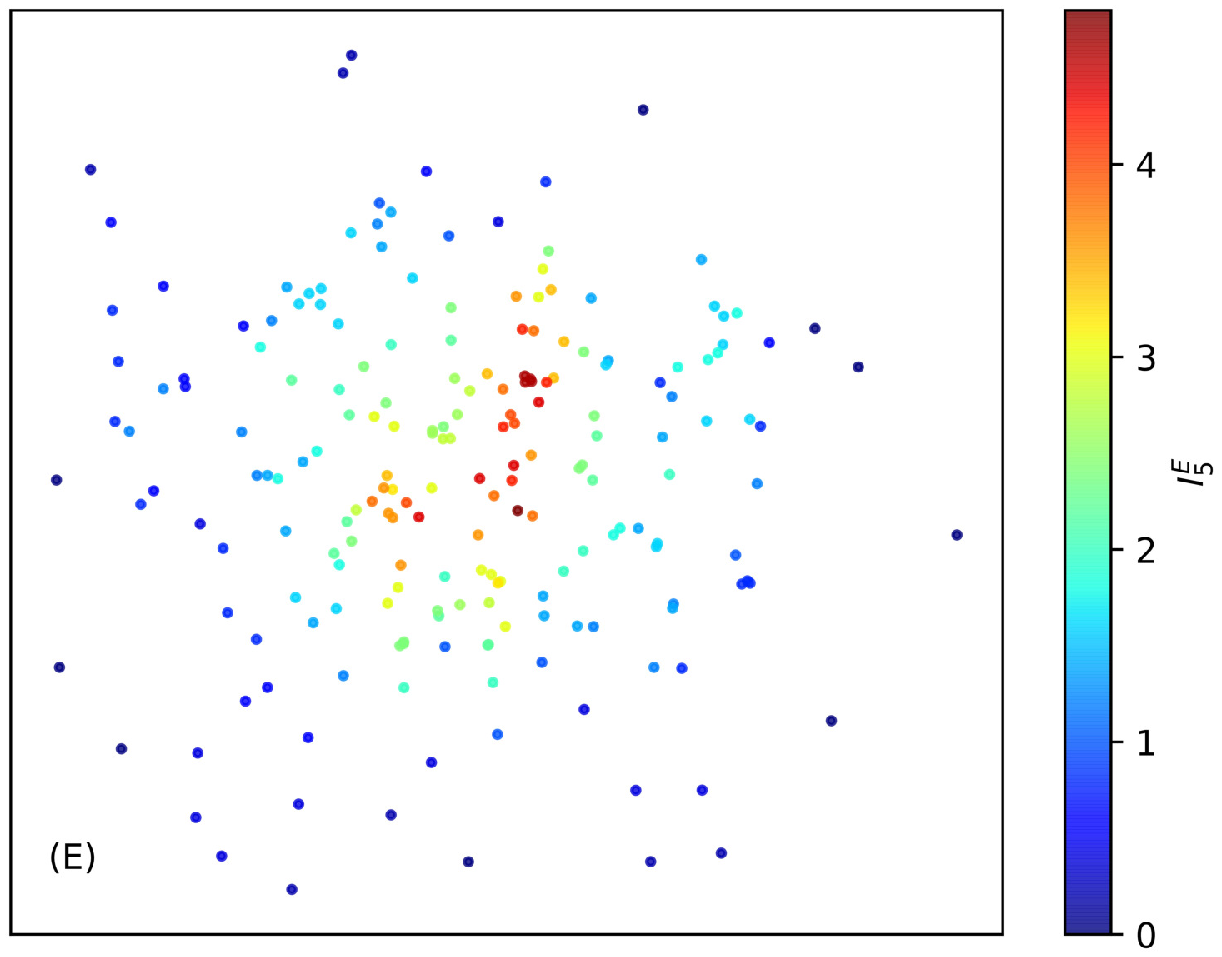} \hfill
   \includegraphics[width=6cm,height=5cm]{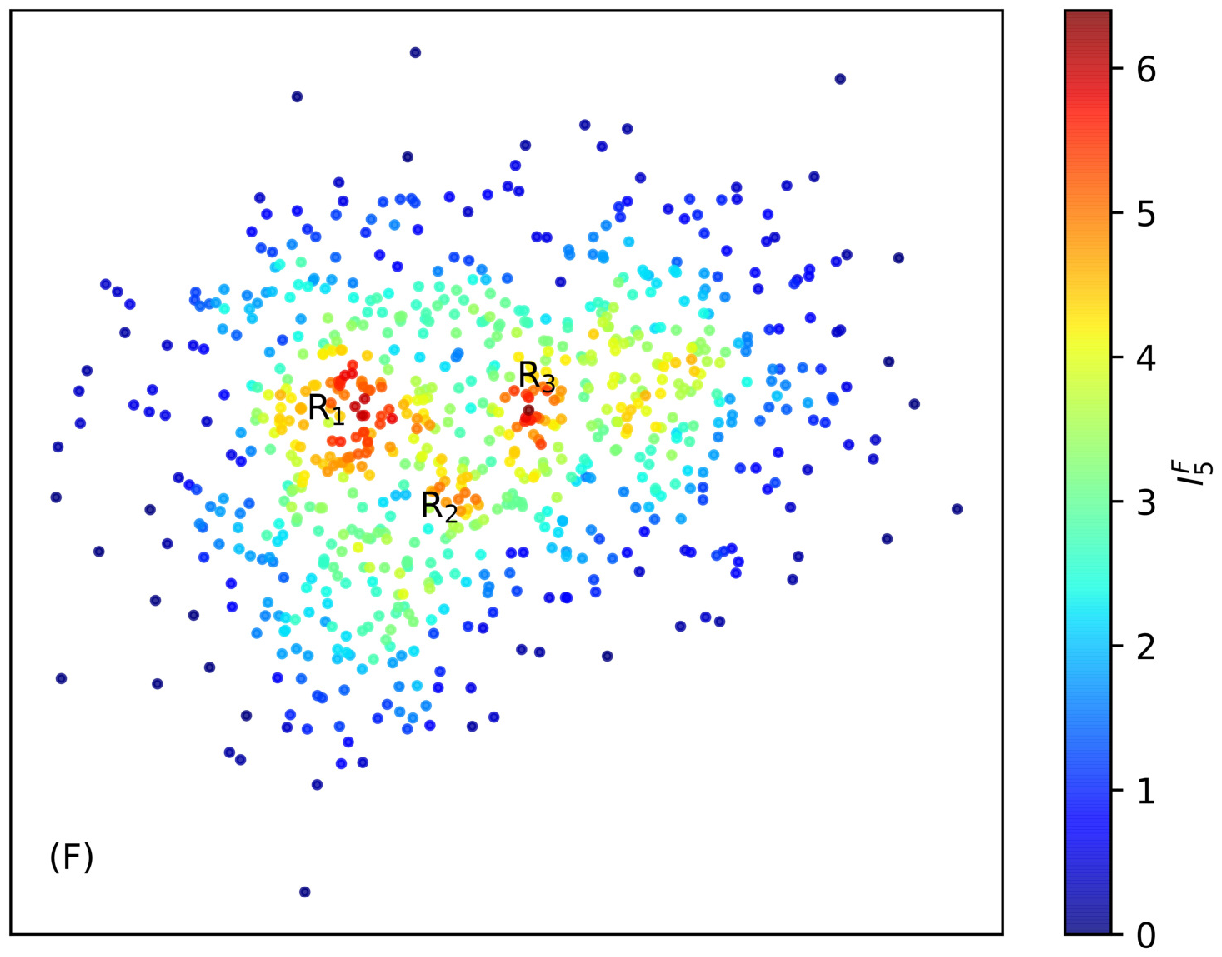} \hfill
  \caption{Plots show the index values derived by INDICATE for members of synthetic clusters D, E, F using a nearest neighbour number of $\mathit{N}=5$ and a standard control distribution (CDA - see Appendix \ref{sec_edge}).}  \label{fig_impl}
\end{figure*}

We conduct a series of statistical tests on randomly generated samples to calibrate, and investigate edge and field effects on, our index. The methodology and results of these tests are described in detail in Appendices \ref{sec_cali}, \ref{sec_edge} and \ref{sec_inter} respectively. In brief:

\begin{enumerate}

\item There is a logarithmic relationship between the maximum index value for a random distribution and sample size. \newline

\item The index is independent of a samples number density. \newline

\item There is a relationship between the typical index value for a random distribution and the chosen N$^{th}$ nearest neighbour number. \newline

\item The size of the control distribution is essentially arbitrary (but care should be taken when a point in the sample has an index value which is on the boundary of a chosen significance threshold). \newline

\item Uniformly distributed interloping field stars (e.g. in observational datasets) typically do not significantly affect the index values of true cluster members.\newline

\item If interloping field stars are distributed in a gradient, the index derived for true cluster members is independent of gradient shape for small nearest neighbour numbers ($\mathit{ N}=3$). \newline

\end{enumerate}

We note that in samples which contain interloping field stars, that the field stars are also assigned index values by INDICATE, so care must be taken when interpreting the values and drawing conclusions on the physical origins of the clustering tendencies of stars in these samples.

%~~~~~~~~~~~~~~~~~~~~~~~~~~~~~~~~~~~~~~~~~~~~~~~~~~~~~~~~~~~~~~~~~~~~

\begin{table}
\caption{Statistics of index values derived for clusters' D, E, F (Fig.\,\ref{fig_impl}). The percentage of members clustered above random ($I_{5}> I_{max}$) is given for each cluster. \label{table_impl}}              % title of Table
\label{table_example}      % is used to refer this table in the text
\centering                                      % used for centering table
\begin{tabular}{c c c c}          % centered columns (4 columns)
\hline\hline                        % inserts double horizontal lines                                   % inserts single horizontal line
Cluster &  $Mo(I_{5})$ & $max(I_{5})$ & $\% \text{ stars }I_{5}> I_{max}$ \\ 
\hline 
D & 8.2 & 10.6 & 81.2 \\
E & 1.6 & 4.8 & 17.5 \\
F & 3.0 & 6.4 & 34.1 \\
\hline 

\end{tabular}
\end{table}

\subsection{Implementation Example}\label{sec_impexample}

In Section\,\ref{sec_obs} we apply our tool on a real stellar catalogue of NGC\,3372. Here we demonstrate using synthetic datasets INDICATE’s ability to quantify the degree of association for each point in a 2D discrete dataset and suitability as a statistical measure for comparative analysis of the spatial distributions of points in multiple datasets. 

Figure\,\ref{fig_impl} shows three clusters (D, E and F) with different degrees of elongation, angular dispersion, sub-structure, surface density and number of members. We apply INDICATE to each using a N$^{th}$ nearest neighbour number of $\mathit{N}=5$ and a standard control distribution (CDA - see Appendix \ref{sec_edge}).  Table\,\ref{table_impl} shows their mode and maximum index values, and the percentage of members with $I_5 > I_{max}$ (Eq.\,\ref{eq_Imax}). 

Cluster D is the most elongated, and its members have been identified by INDICATE as having the strongest clustering tendencies (highest index values) of the three clusters. We find that the greatest degree of association is within its central region (up to 53 neighbours within $\bar{r}$) and spatial clustering of members is asymmetrical – stars to the NE of the highest index members have significantly higher index values than those to the SW. 

Members of cluster E have a markedly different spatial distribution and clustering tendencies to those of cluster D. The spatial distribution of members lacks a strong radial correlation and two concentrations of (relatively) high index stars are identified. Stars with the greatest degree of association have 24 neighbours within $\bar{r}$ (a factor of 2 less than D), typically the degree of association of members of E is a factor of 6.6 less than members of cluster D (i.e. members in E are significantly less tightly clustered than those of D). Of course, as the clusters are being analysed in a 2D parameter space it is conceivable that stars in cluster D and E may have very similar spatial distributions but are being viewed from different rotations around the 3D axis. We therefore advise caution when drawing conclusions from comparisons of the INDICATE values of two or more observational 2D datasets alone. 

Cluster F has three concentrations of high index members (marked on Fig.\,\ref{fig_impl} as $R_1$, $R_2$, $R_3$). Stars that form part of $R_1$ and $R_3$ have similar index values, that is they have similar clustering tendencies, but those of $R_2$ are less spatially clustered.  If this were a real dataset,  where clustering behaviours are dictated by underlying physics, these index values could be used as a starting point to  explore the physical causes of the identified discrepancies between the clustering behaviours of the three concentrations (e.g. differences in evolutionary stage, initial conditions, stellar mass of members) and also the identified disparity of members clustering behaviours between clusters D, E, and F. This form of quantitative analysis and comparison of spatial behaviours, as achieved here by INDICATE, is not possible with the discussed global methods and/or established clustering algorithms.

%######################################################################################################
\section{Tracing Morphological Features}\label{sec_obs}

%######################

The Carina Nebula (NGC\,3372)  is a massive star forming HII complex in the southern sky at a distance of 2.3kpc, containing $>10^{5}\,M_{\odot}$ of gas+dust \citep{2011A&A...525A..92P}. It is one of the nearest and richest concentrations of OB stars ($>$130; \citealt{2014ApJ...787..107K}) in the Galaxy and includes some of the most massive and luminous known single and binary stars (e.g. Eta Carinae, HD 93129, W25). The region is well studied and has considerable sub-structure including the Trumpler 14-16, Collinder 228, Collinder 232, Collinder 234, Bochum 10 and Bochum 11 clusters. Triggered star formation is ongoing in the complex, driven by massive star feedback (\citealt{2008hsf2.book..138S}; \citealt{2010MNRAS.406..952S}; \citealt{2011A&A...530A..34P}; \citealt{2013A&A...549A..67G}). Coupled with its low line of sight extinction, the Carina Nebula is therefore an ideal laboratory in which to study massive star formation. For a review of the region see \citet{2008hsf2.book..138S}.

\begin{figure*}
\centering
   \includegraphics[width=0.8\textwidth]{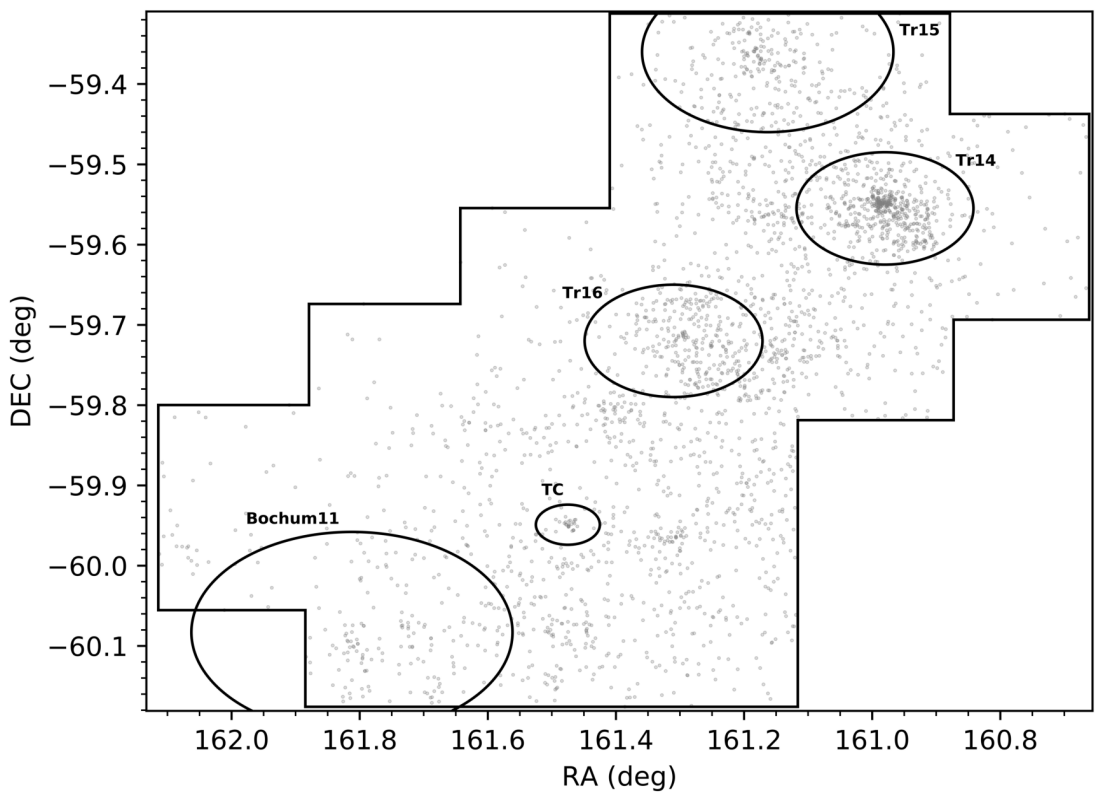} 
  \caption{Plot of the stellar catalogue by \citet{2014ApJ...787..107K}, with stars represented as grey dots. The positions of the Trumpler\,14 (Tr14), Trumpler\,15 (Tr15), Trumpler\,16 (Tr16), Treasure Chest (TC) and Bochum 11 clusters are overlaid as black ellipses (see text for details).}  \label{fig_carina_catalogue}
\end{figure*}

\begin{table}
\caption{Statistics for stars within the radial boundaries of the Tr14, Tr15, Tr16, TC and Bochum\,11 clusters with an index value of $I_5>I_{sig}$. \label{tab_carina}}              % title of Table
\label{table_example}      % is used to refer this table in the text
\centering                                      % used for centering table
\begin{tabular}{c c c c}          % centered columns (4 columns)
\hline\hline                        % inserts double horizontal lines                                   % inserts single horizontal line
Cluster &  Total & $\bar{I_5}$ & $\max I_5$  \\ 
\hline 
Tr14 &  470 (85.2$\%$)& 12.4 & 29.6  \\ 
Tr15 & 75 (29.1$\%$) &  3.9 & 6.6  \\ 
Tr16 & 212 (73.4$\%$) &  3.6 & 5.6  \\ 
TC & 25 (83.3$\%$) & 4.3 & 4.8  \\ 
Bochum\,11 & 10 (5.8$\%$) & 2.7 & 3.0  \\ 
\end{tabular}
\end{table}

\begin{figure*}
\centering
   \includegraphics[width=0.8\textwidth]{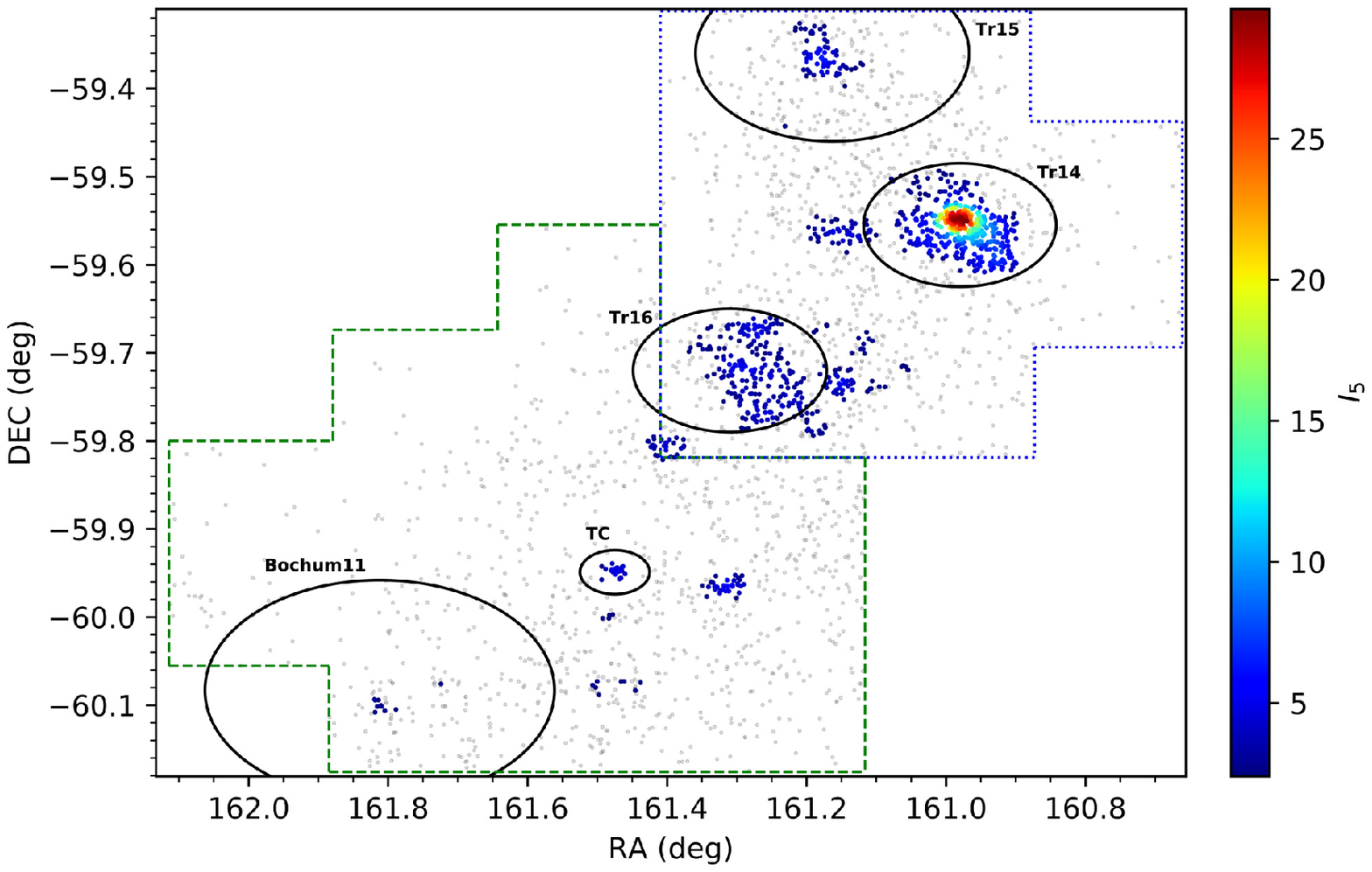} 
   \includegraphics[width=0.8\textwidth]{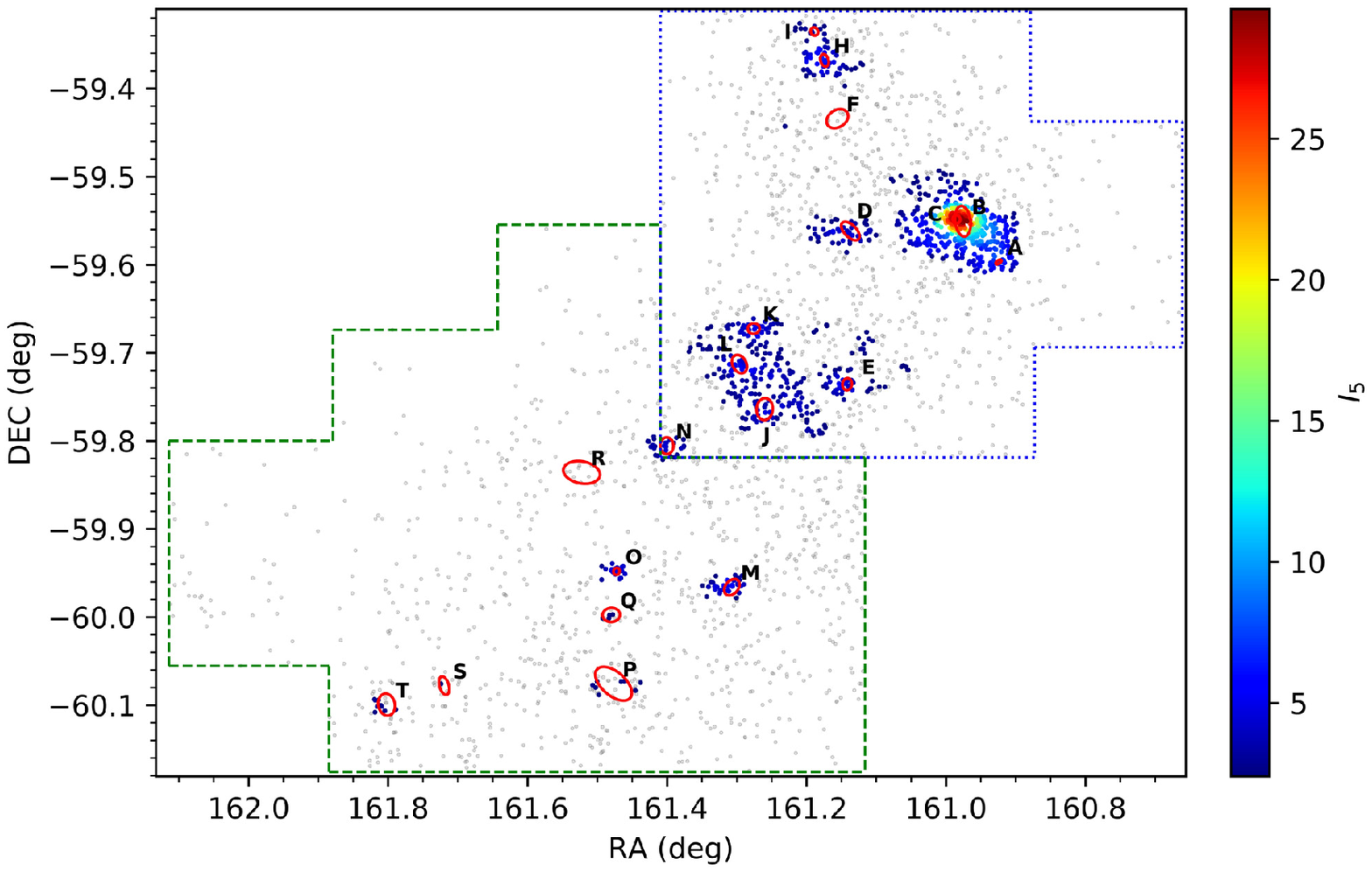} 
  \caption{Plots of index values, $I_{5}$, calculated by INDICATE for the Carina region with positions of the (Top:) Tr14, Tr15, Tr16, TC and Bochum\,11 clusters overlaid as black ellipses and (Bottom:) 19 sub-clusters identified by \citet{2014ApJ...787..107K} overlaid as red ellipses (see text for details). The borders of our designated NW and SE regions are marked with blue dotted and green dashed lines respectively. Stars with an index value above the significance threshold ($I_{5}>2.3$) coloured as described by the colour bar. Grey dots are stars with $I_{5}<2.3$.}  \label{fig_carina_knownclusters}
\end{figure*}

\begin{figure*}
\centering
   \includegraphics[width=0.8\textwidth]{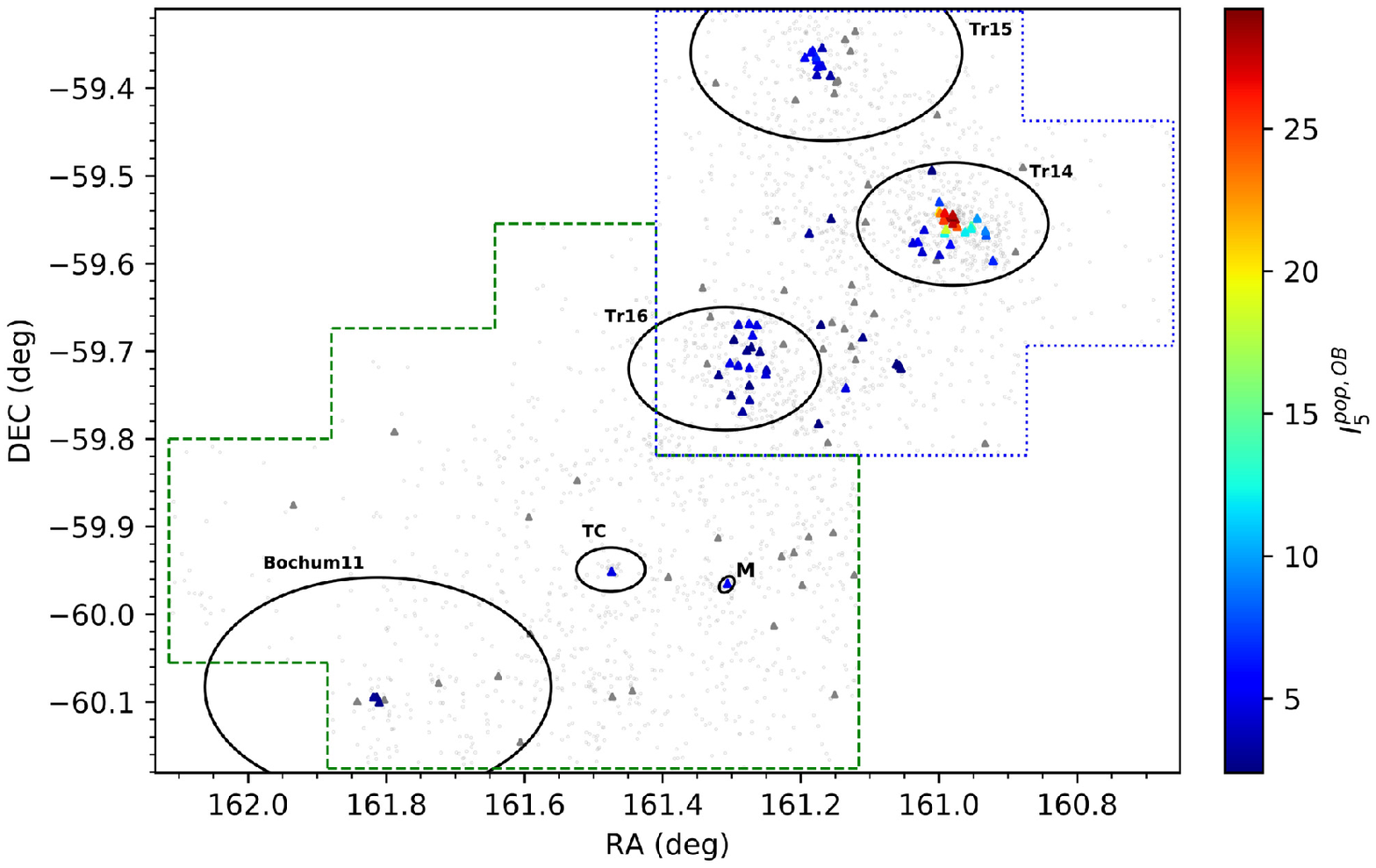} 
   \includegraphics[width=0.8\textwidth]{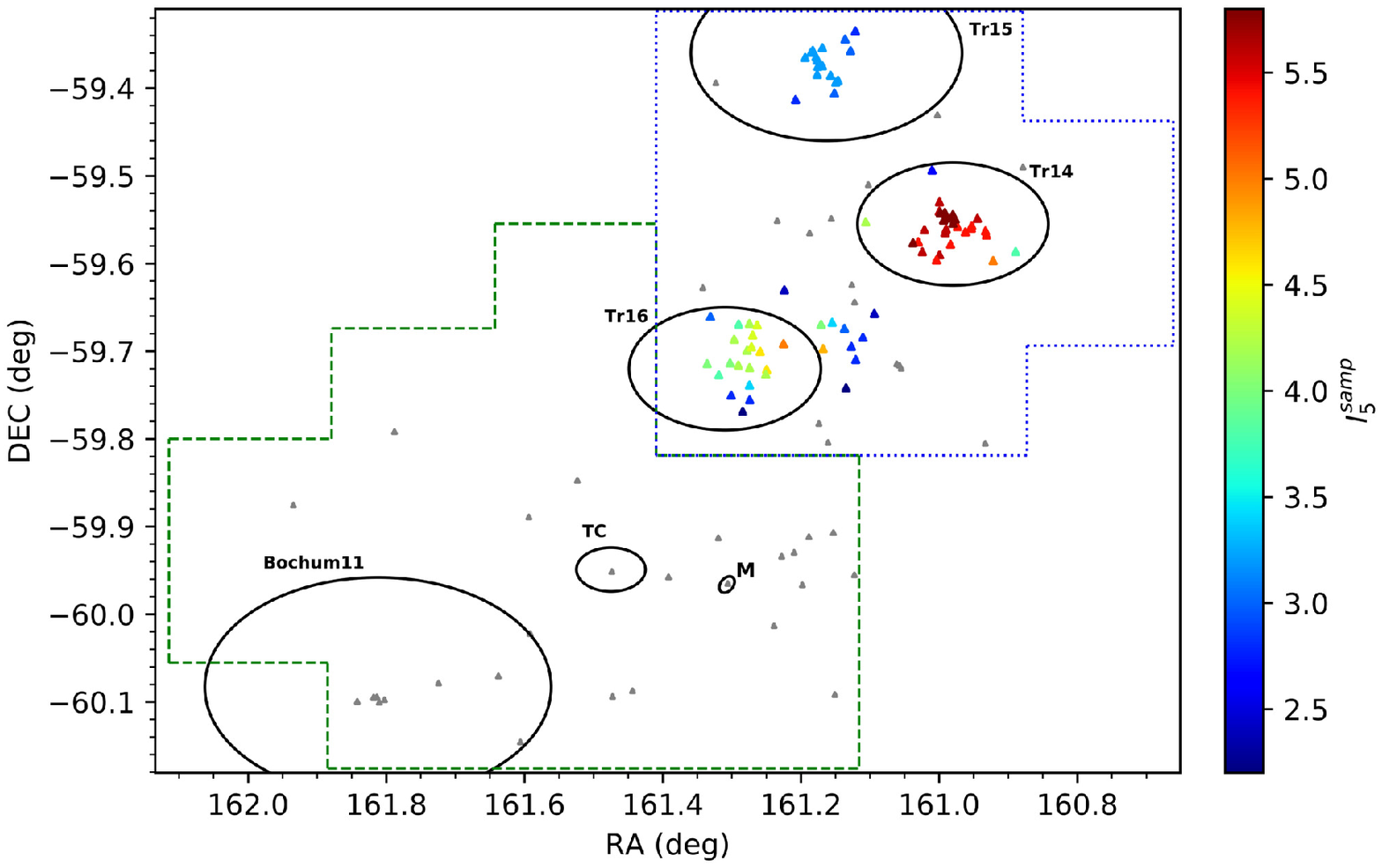} 
  \caption{Plots of index values calculated for the OB population by INDICATE when applied to (Top:) the entire stellar catalogue, $I^{pop, OB}_{5}$; and (Bottom:) OB sub-sample, $I^{samp}_{5}$ (see text for details). Positions of the Tr14, Tr15, Tr16, TC, Bochum\,11 and M (sub-) clusters are overlaid as black ellipses. The borders of our designated NW and SE regions are marked with blue dotted and green dashed lines respectively. Coloured triangles represent OB stars with an index value above the respective significance thresholds. Grey triangles are OB stars below the respective significance thresholds.}  \label{fig_carina_OB}
\end{figure*}

\begin{figure*}
\centering
   \includegraphics[width=0.3\textwidth]{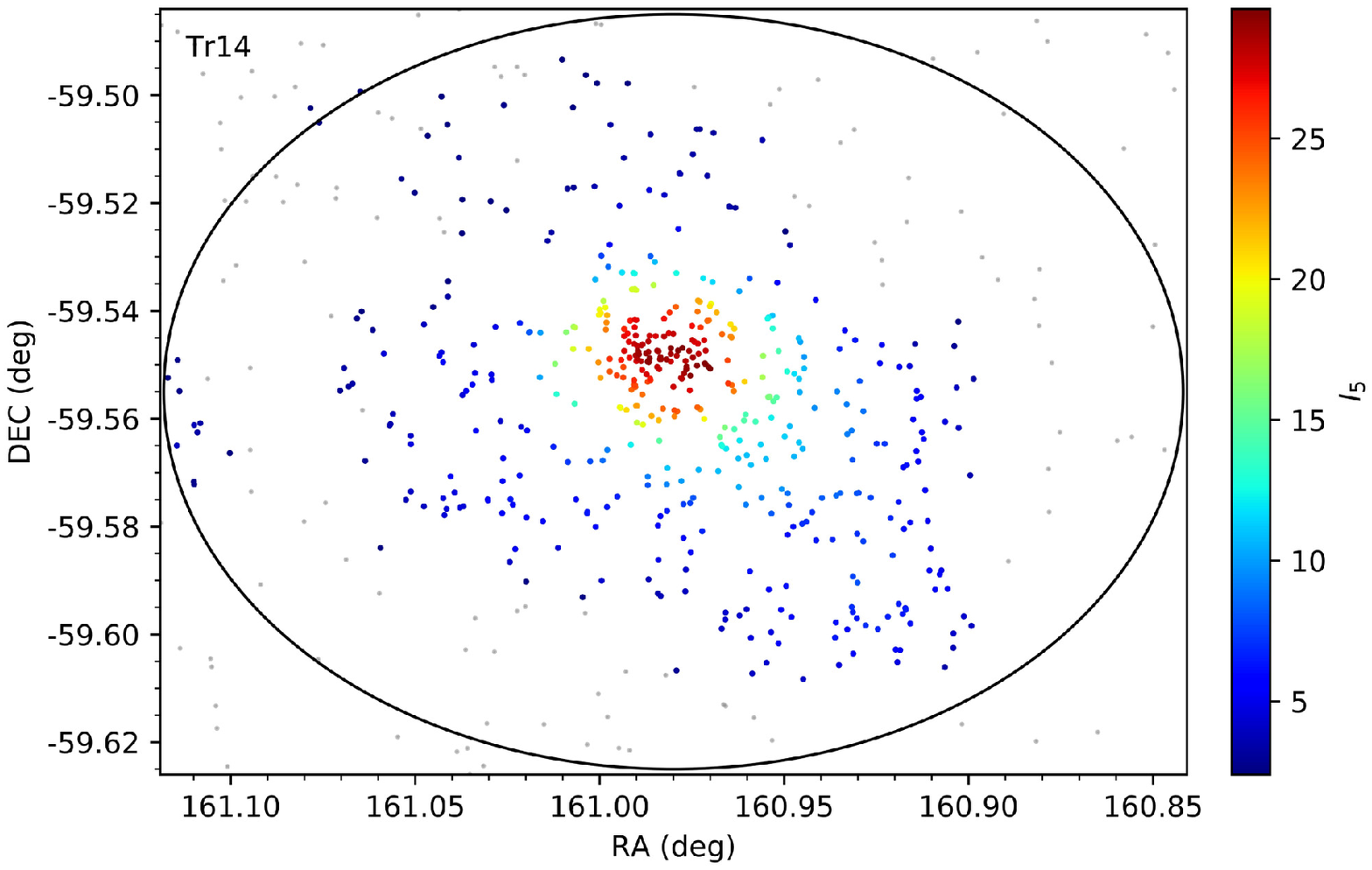} 
   \includegraphics[width=0.3\textwidth]{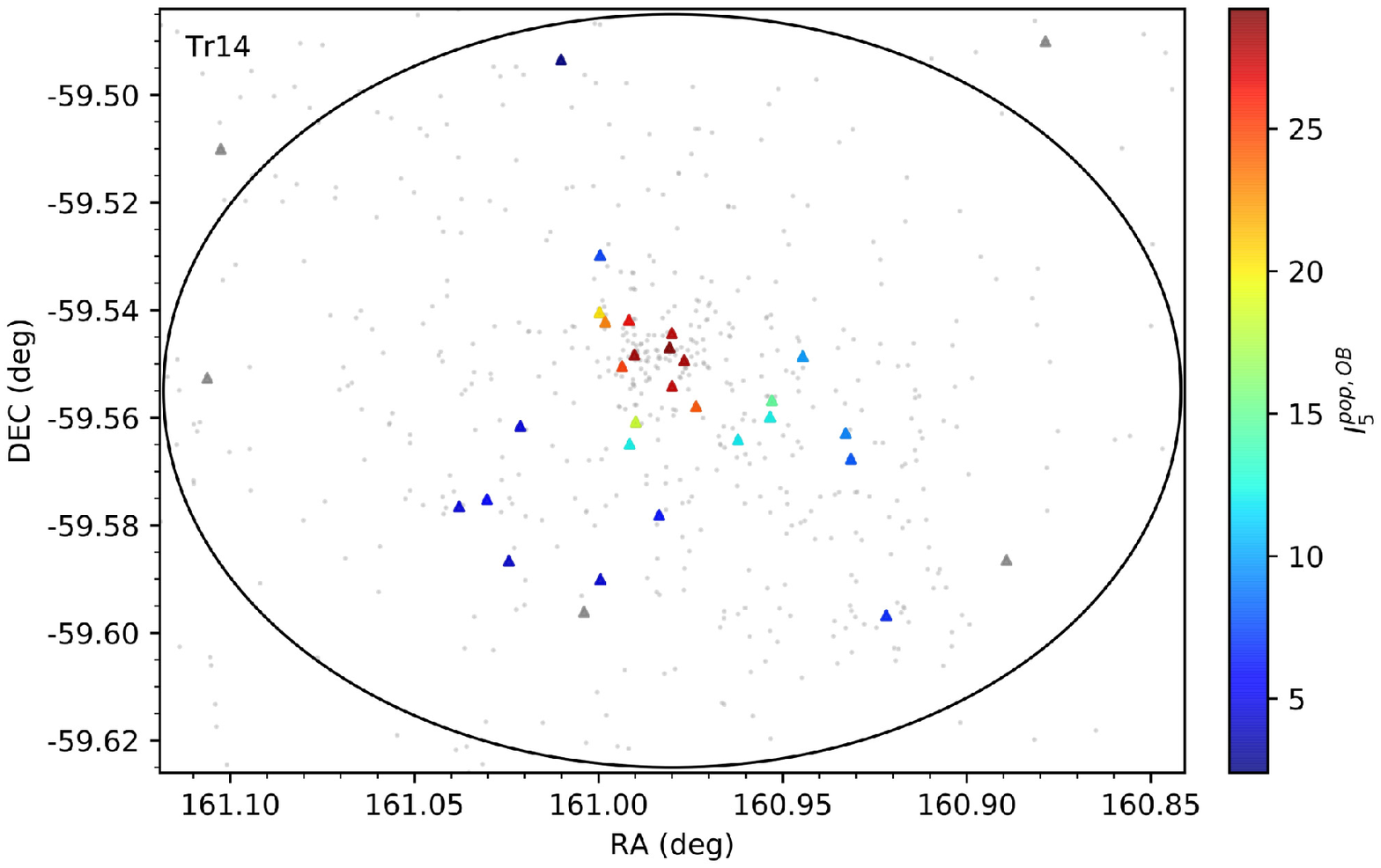} 
   \includegraphics[width=0.3\textwidth]{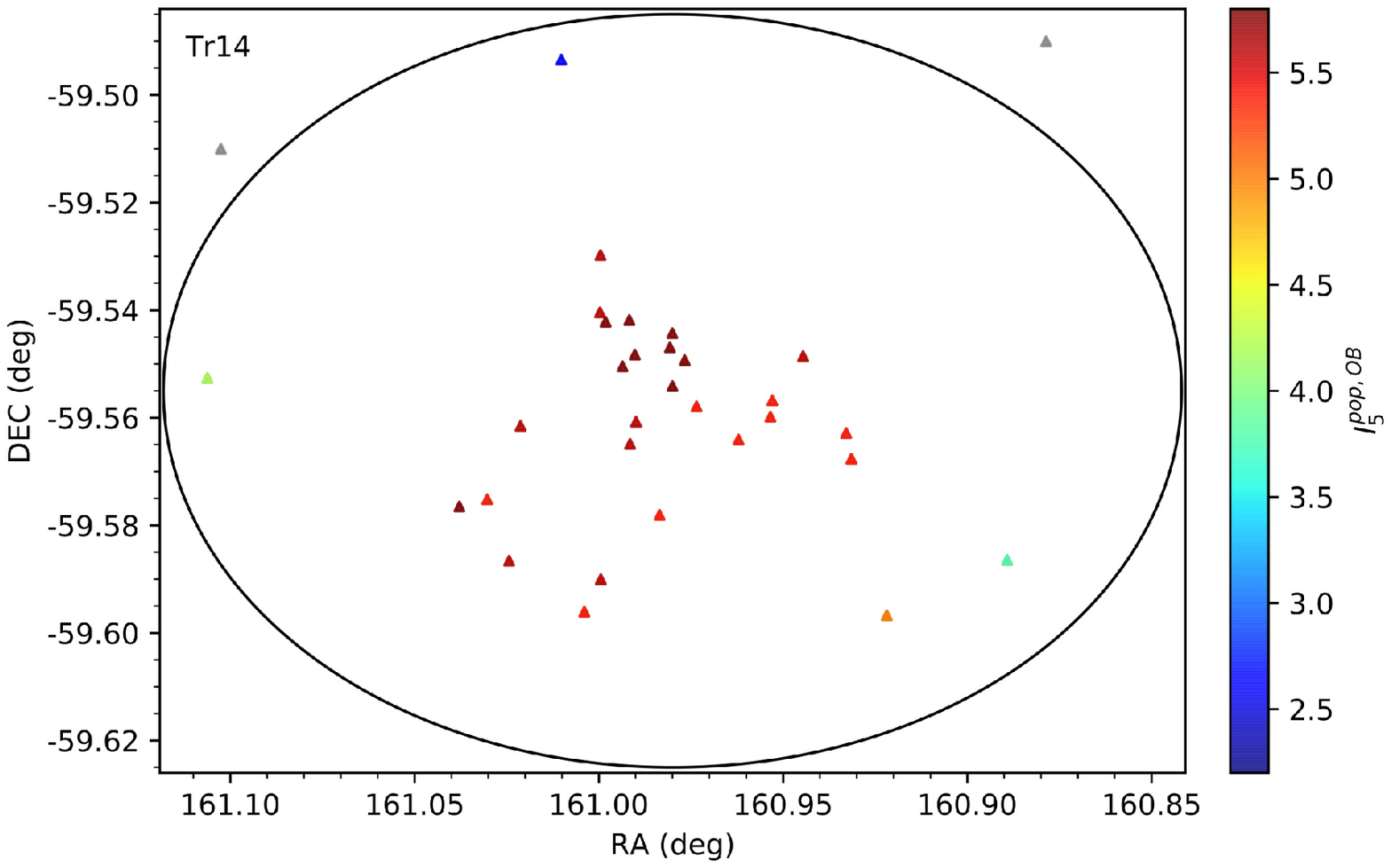} 

   \includegraphics[width=0.3\textwidth]{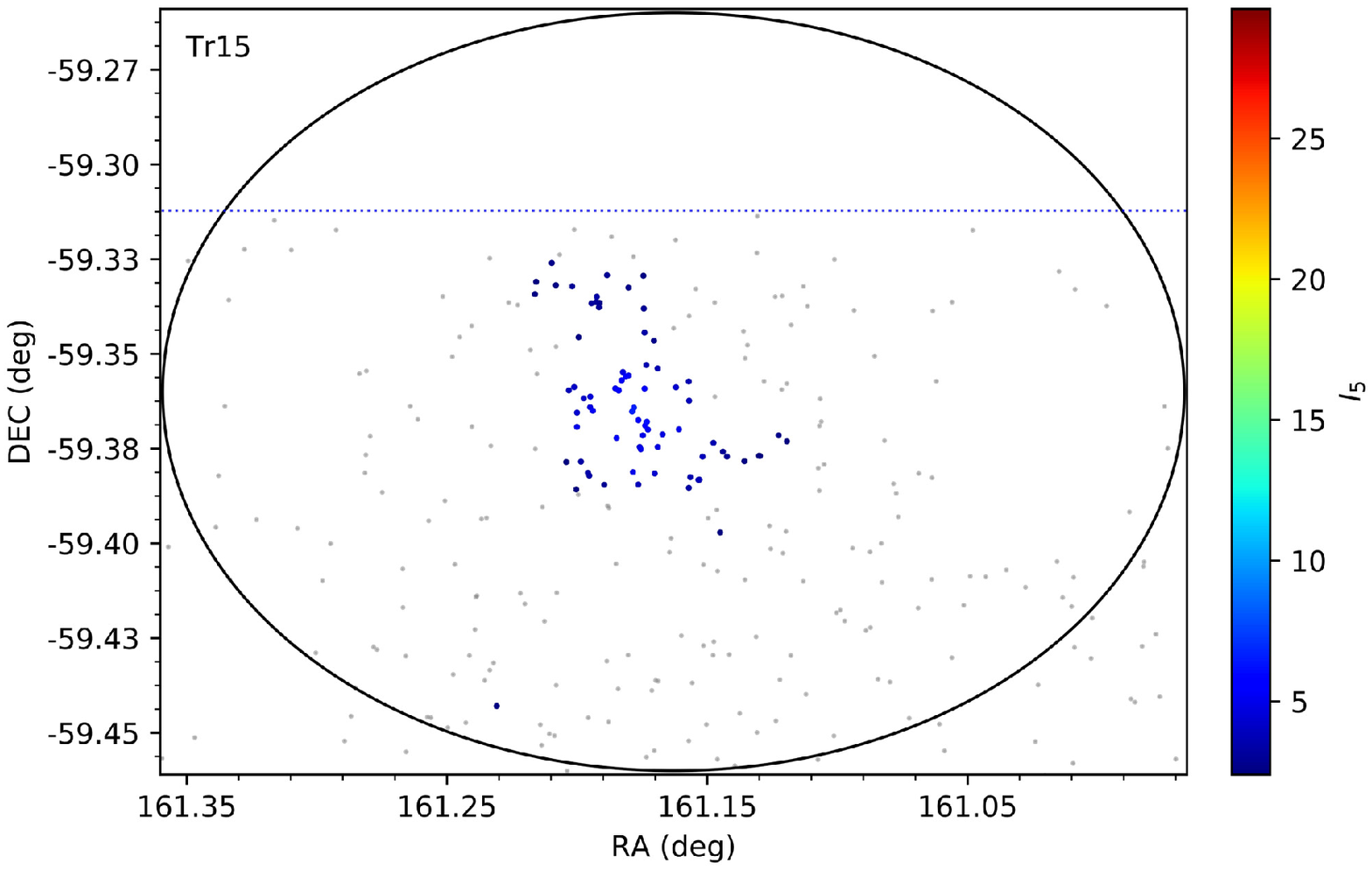} 
   \includegraphics[width=0.3\textwidth]{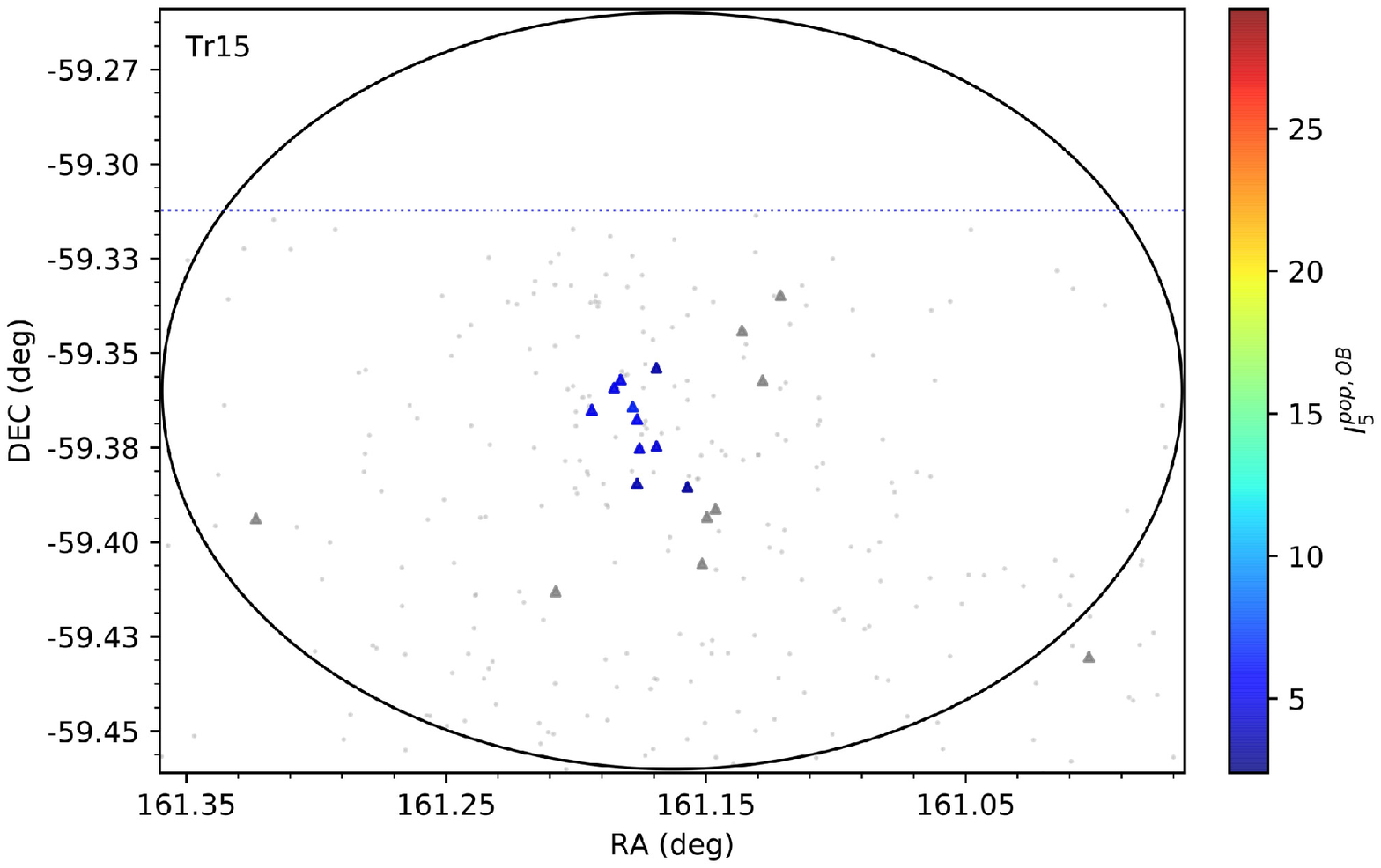} 
   \includegraphics[width=0.3\textwidth]{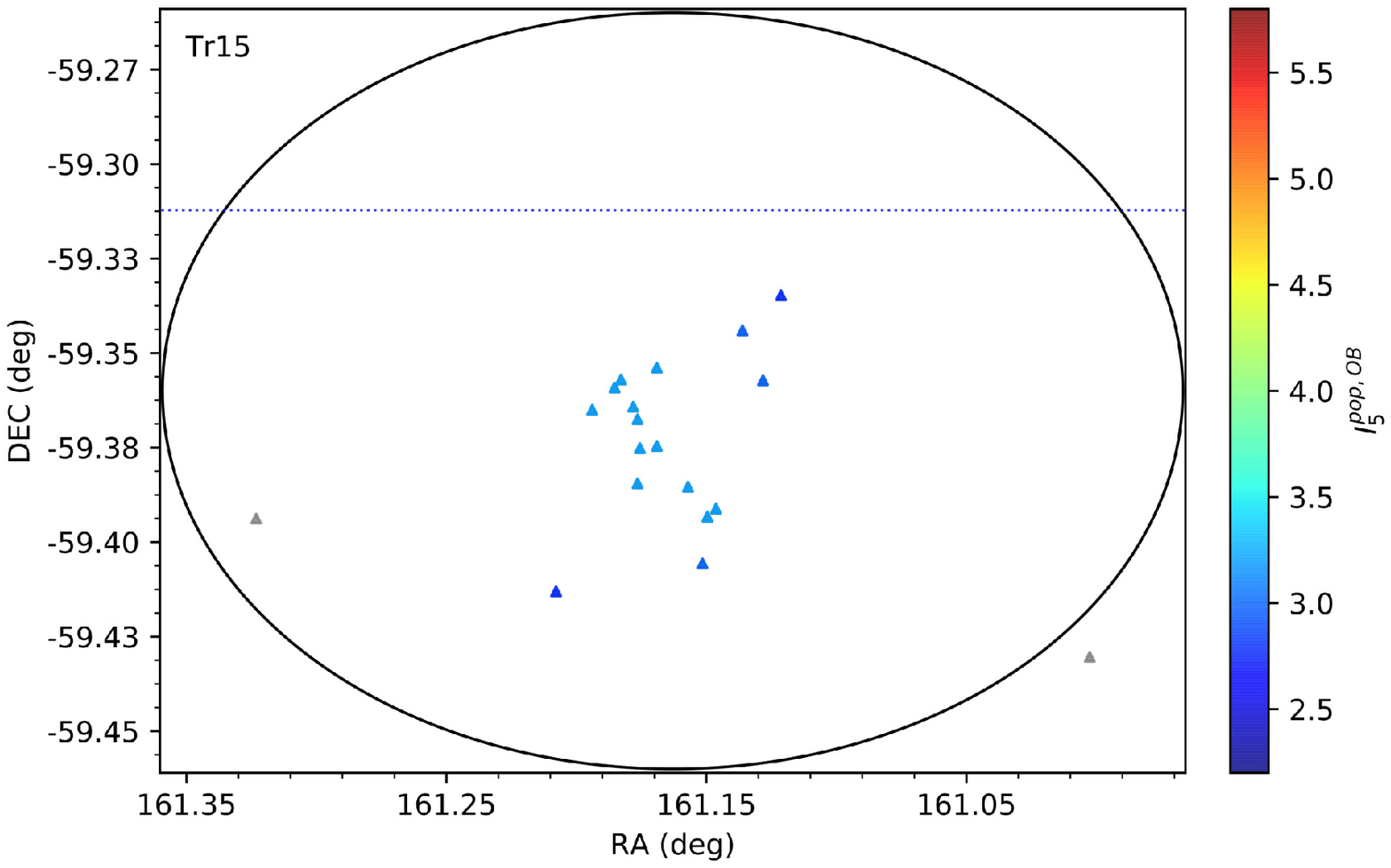} 

   \includegraphics[width=0.3\textwidth]{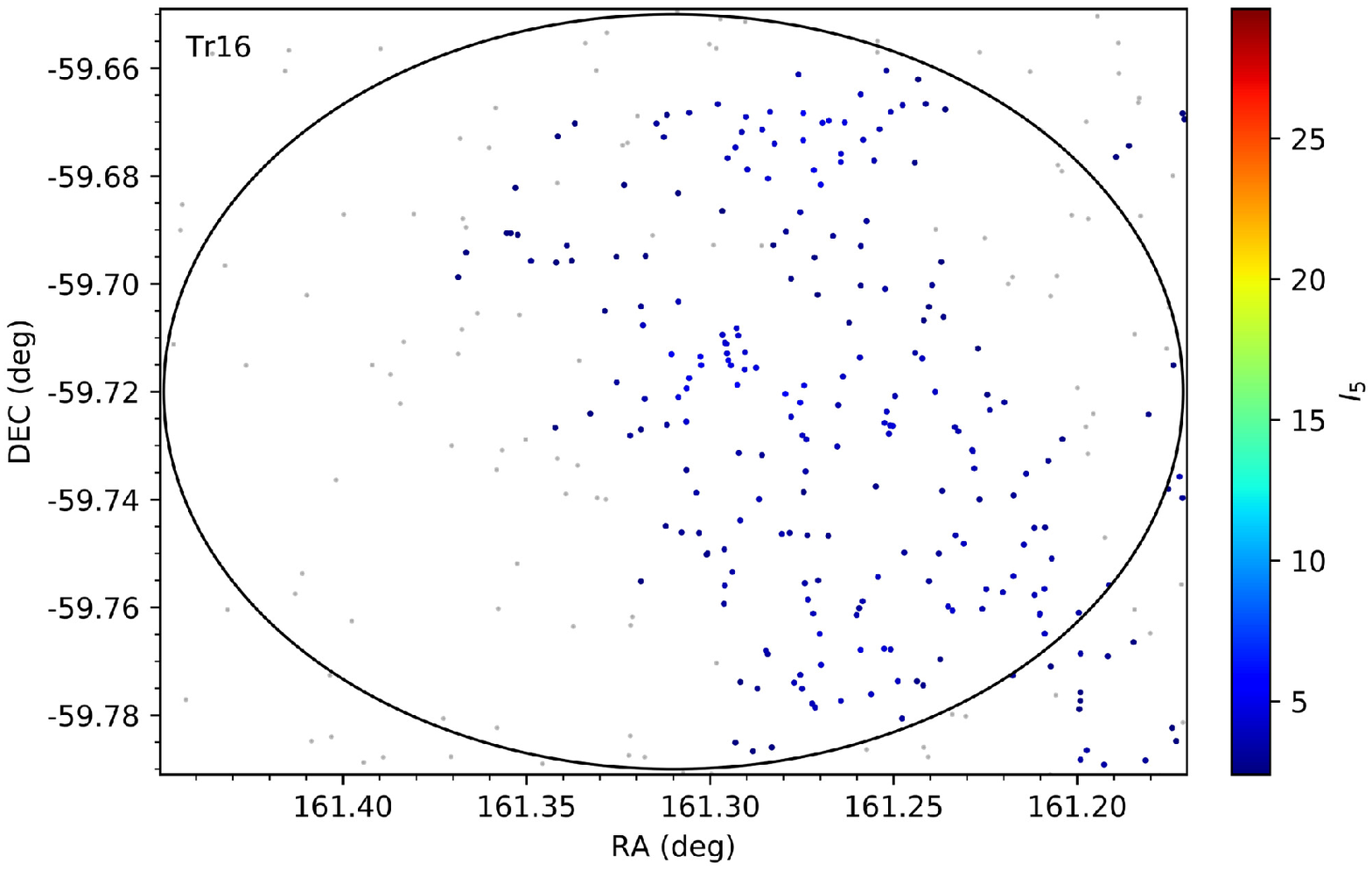} 
   \includegraphics[width=0.3\textwidth]{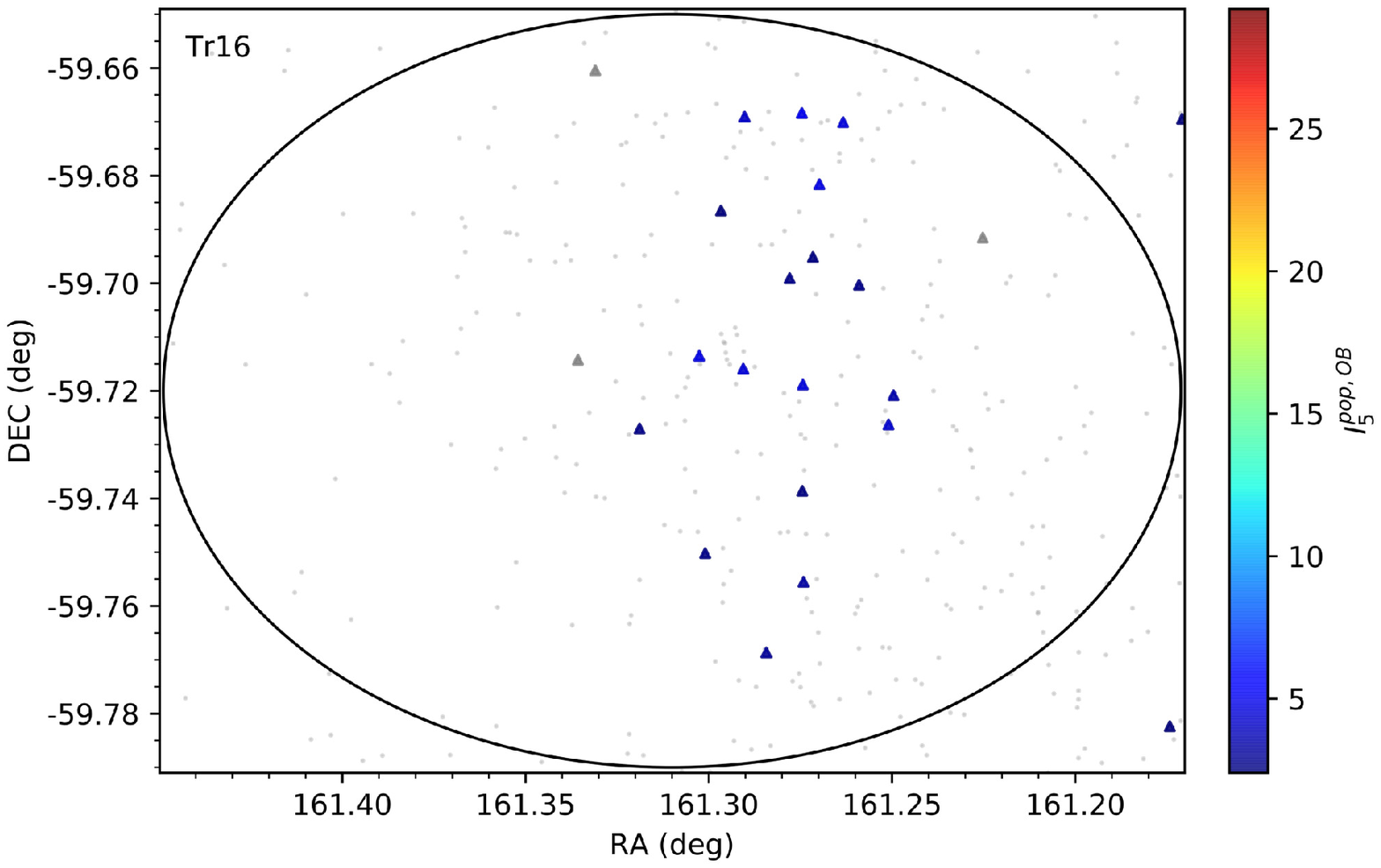} 
   \includegraphics[width=0.3\textwidth]{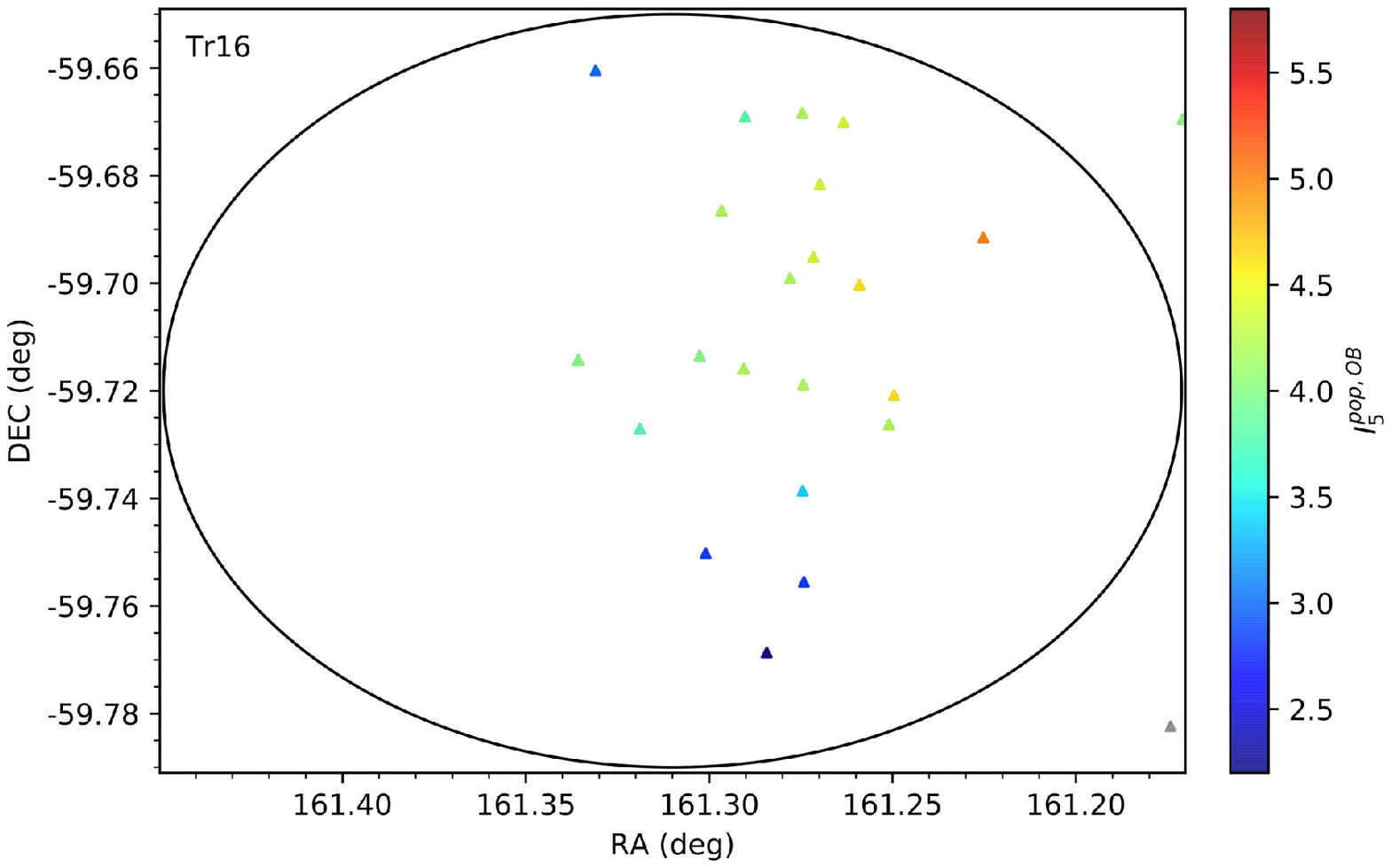} 

   \includegraphics[width=0.3\textwidth]{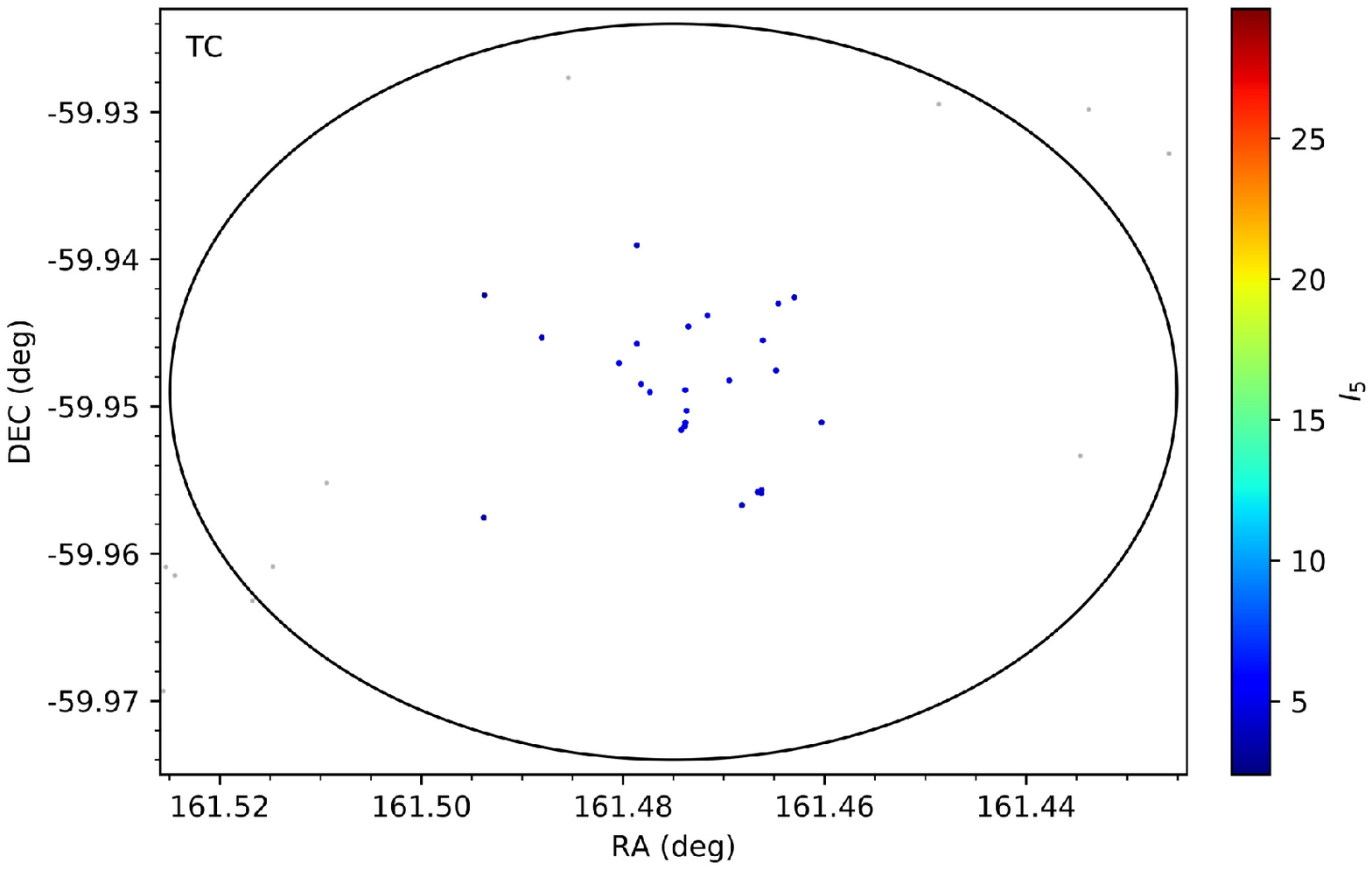} 
   \includegraphics[width=0.3\textwidth]{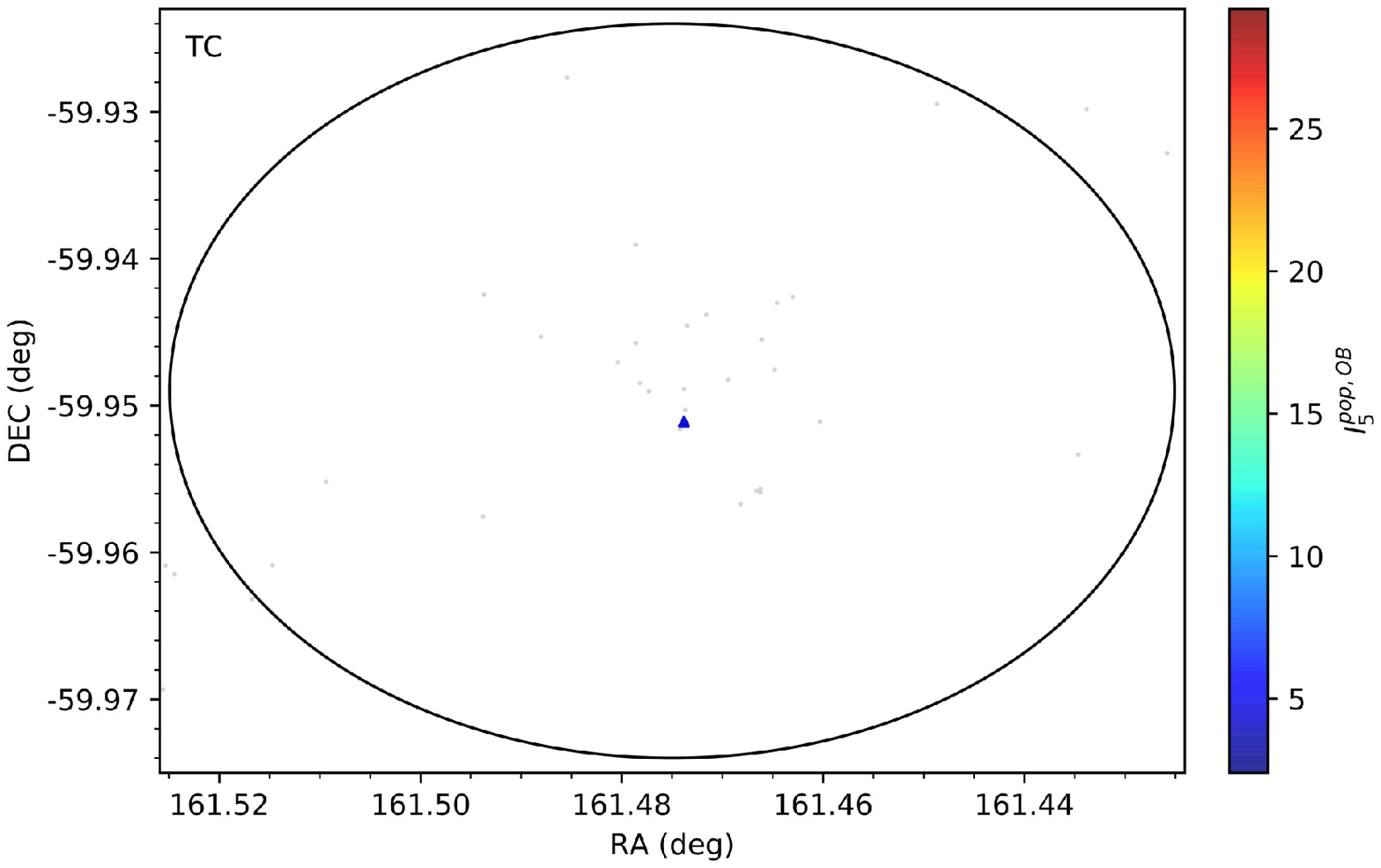} 
   \includegraphics[width=0.3\textwidth]{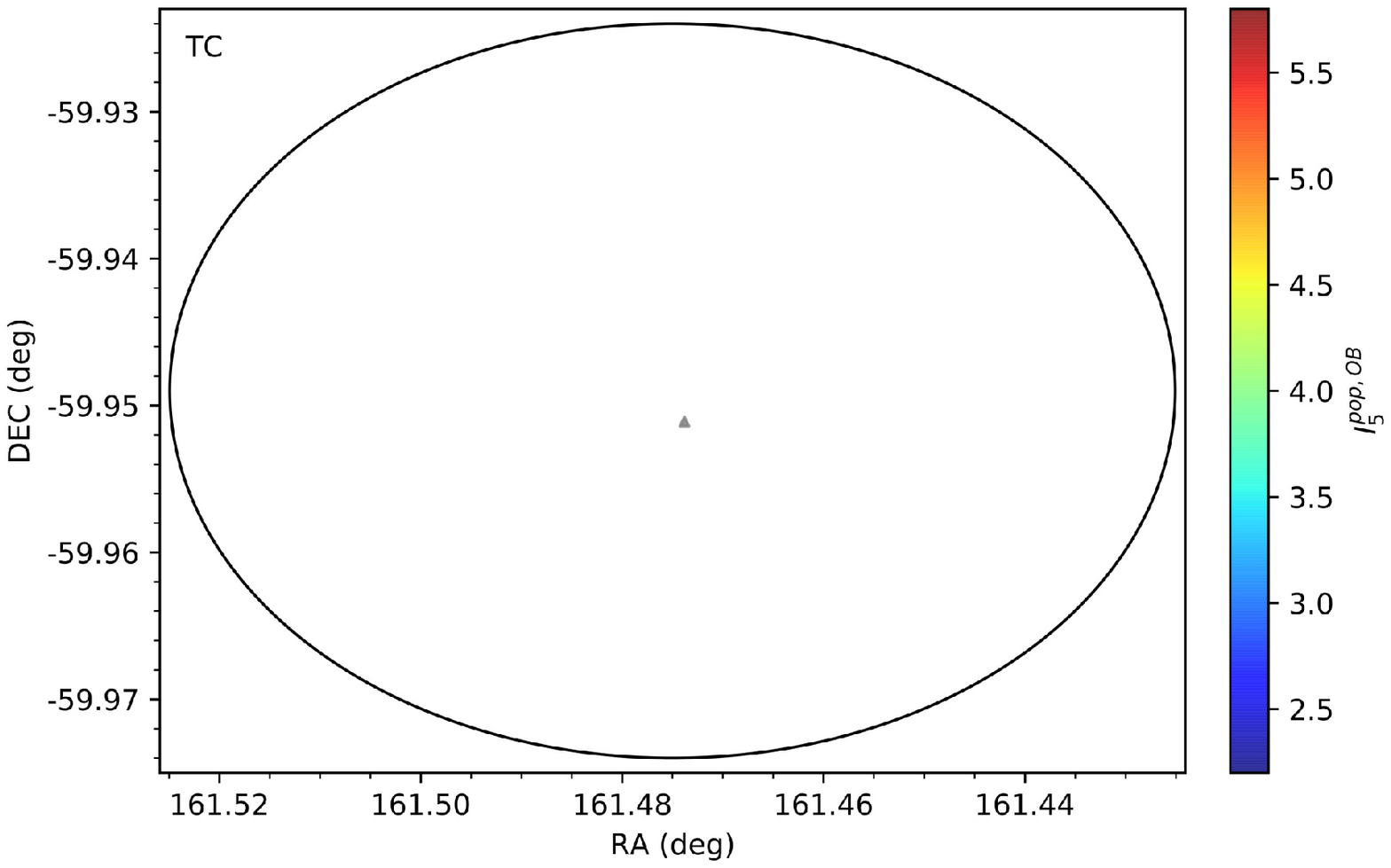} 

   \includegraphics[width=0.3\textwidth]{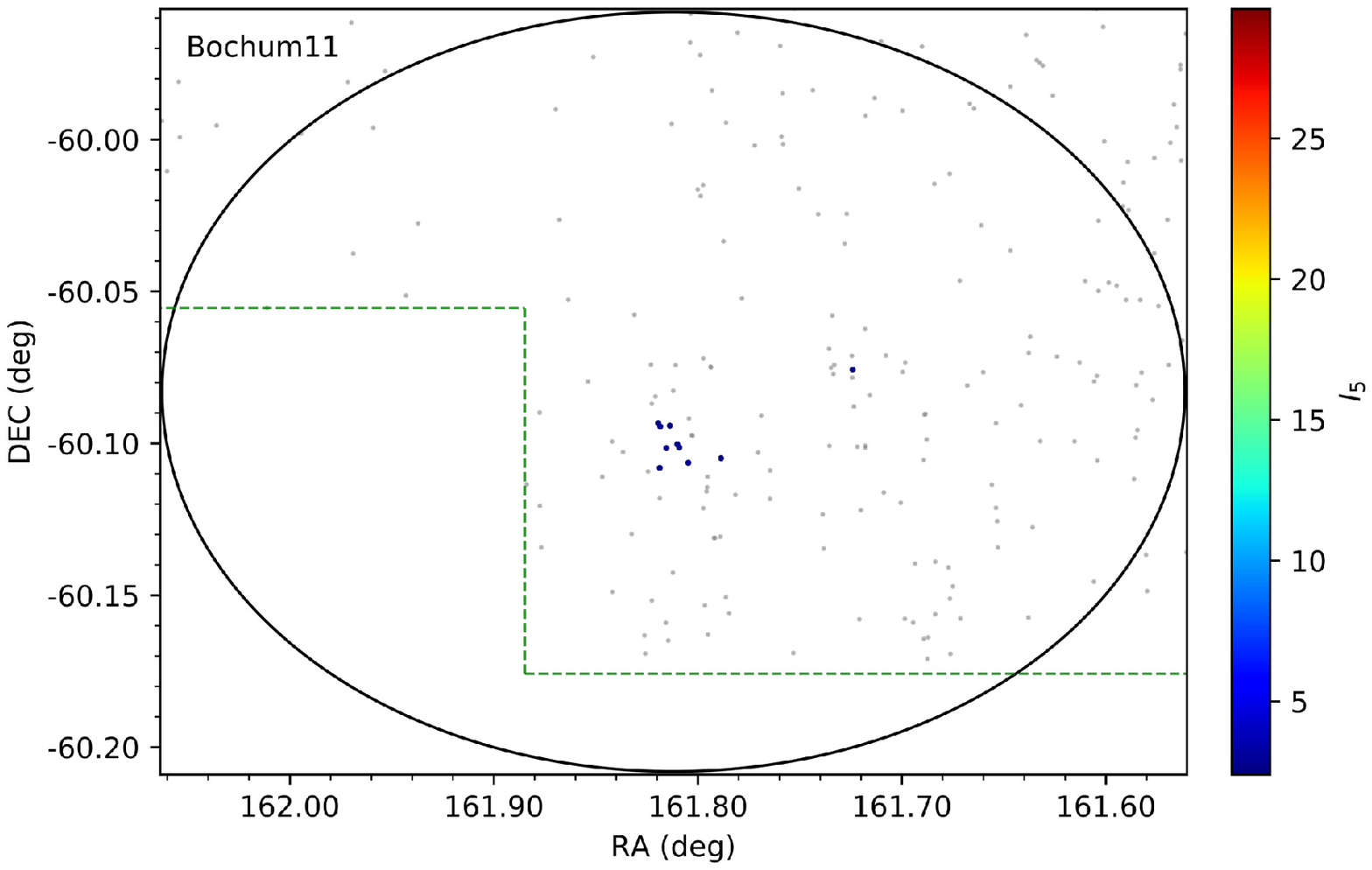} 
   \includegraphics[width=0.3\textwidth]{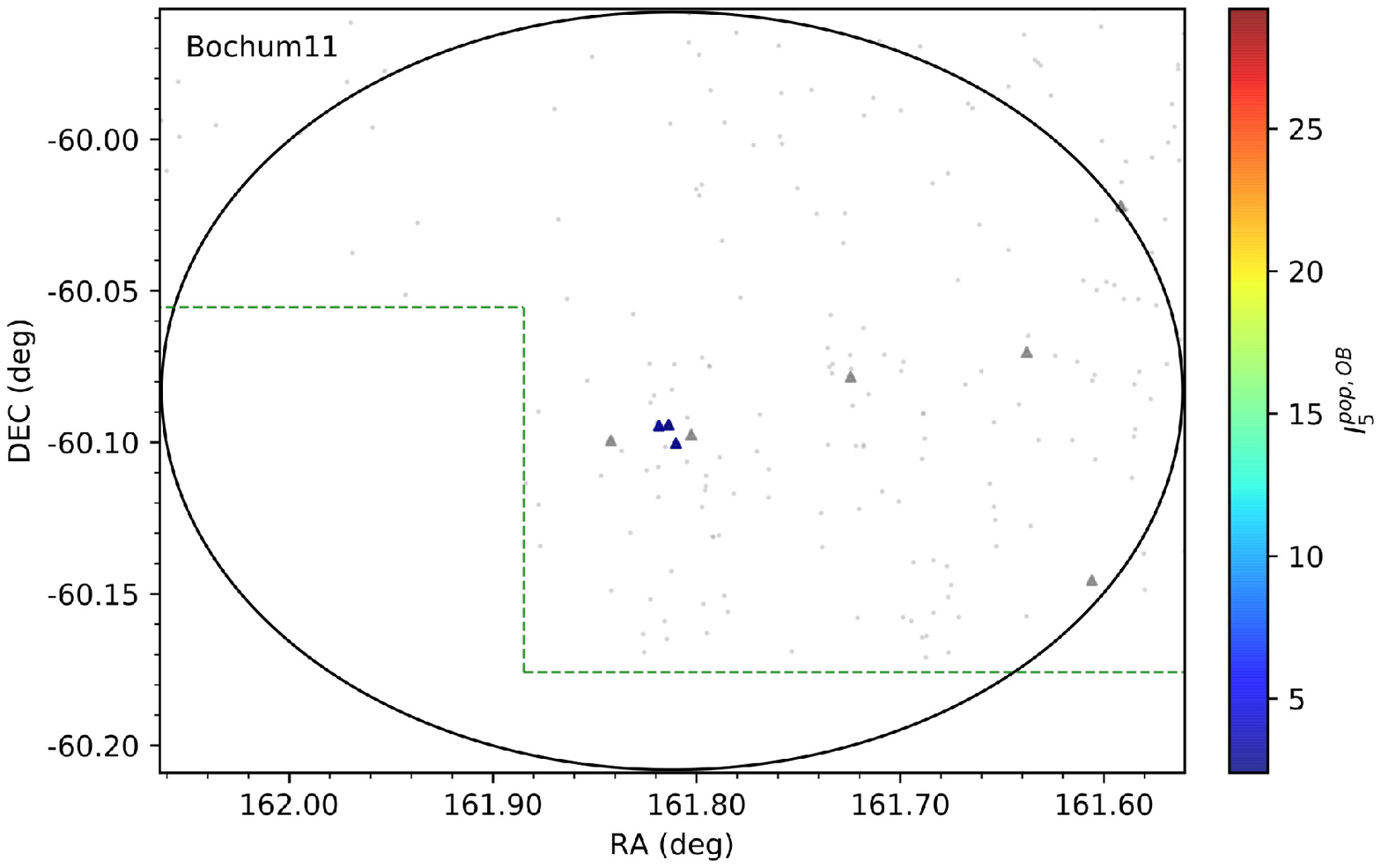} 
   \includegraphics[width=0.3\textwidth]{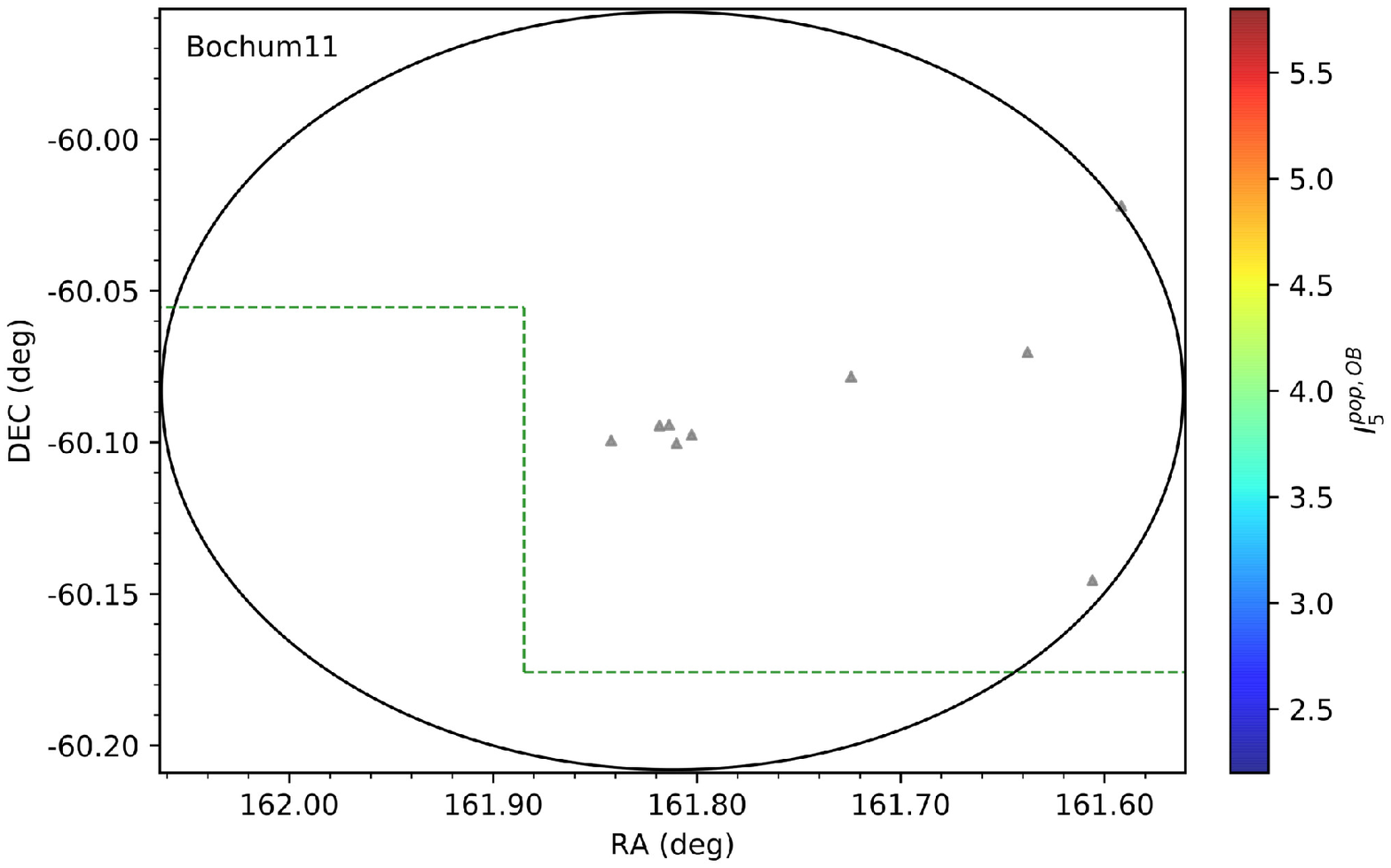} 

  \caption{Zoomed-in plots of clusters Tr14, Tr15, Tr16, TC and Bochum11 as shown in (Left:) top of Fig\,\ref{fig_carina_knownclusters}, (Middle:) top of Fig\,\ref{fig_carina_OB} and (Right:) bottom of Fig\,\ref{fig_carina_OB}.}  \label{fig_carina_zoomin}
\end{figure*}

\subsection{Identification of Stellar Structure}\label{sect_struct}

We apply INDICATE to the stellar catalogue of 2790 stars for the region as described by \citet{2014ApJ...787..107K}, and plotted in Figure\,\ref{fig_carina_catalogue}, with a nearest neighbour number of $\mathit{N}=5$ and an extended control distribution (CDB - see Appendix \ref{sec_edge}). This catalogue was selected because it covers an inner region of the Carina Complex ($\sim\,0.38^{\circ}$) which is rich in sub-structure, containing at least 20 sub-clusters (detected by the original authors) and includes the young Trumpler\,14 (Tr14), Trumpler\,15 (Tr15), Trumpler\,16 (Tr16), Treasure Chest (TC) and Bochum 11 clusters. The position and radius of the TC cluster in Figure\,\ref{fig_carina_catalogue} is as given by \citet{2001A&A...376..434D}, and the position and radius of Tr14-16 and Bochum\,11 clusters by the MWSC catalogue \citep{2013A&A...558A..53K}. On visual inspection, stars in the North West (NW) region of the catalogue have significantly higher degrees of association than those in the South East (SE) region. 

The top plot in Figure\,\ref{fig_carina_knownclusters} and left plots of Fig.\,\ref{fig_carina_zoomin}, show the distribution of stars with their index values and the boundaries of the Tr\,14-16, TC, Bochum\,11 clusters overlaid. We define a significance threshold - that is the value of $I_5$ above which a star is significantly clustered above random - of three standard deviations above the mean value expected from a random distribution of the same size evaluated with  $\mathit{N}=5$ and CDB, such that $I_{sig}=\bar{I}^{random}_5+3\sigma = 2.3$.  All five clusters are clearly identified by stars within their radial boundaries having an index above the defined significance threshold. This is an expected result as by definition the spatial distribution of cluster stars - particularly at their centres - should display a higher degree of clustering than a random (and background) field. 

Table\,\ref{tab_carina} gives statistics on the index values derived for each cluster. More than $80\%$ of the stars within the bounds of Tr14 and TC clusters are clustered above random, which is markedly larger than Tr15 ($29.1\%$) and Bochum 11 ($5.8\%$). Trumpler\,16 has a comparable proportion of stars with index values above the threshold to that of the Tr14 and TC clusters ($73.4\%$) but unlike the other four clusters these stars are not centrally concentrated and instead are in less compact concentrations across the cluster region, which is consistent with results of previous studies of the cluster's structure (e.g. \citealt{2011ApJS..194...12W}). Interestingly, stars clustered above random in Tr15, Tr16 and TC have similar mean index values - that is they have similar degrees of association and clustering tendencies. Stars within the bounds of Tr14 display the highest degree of clustering behaviour with a mean index value a factor of 3 larger than those of the Tr15, Tr16 and TC clusters. In addition, Tr14 also contains the most spatially clustered stars in the Carina region, centrally concentrated at its core, with stars here having up to an additional 137 stars in their local neighbourhoods above that expected in a spatially random distribution. By contrast, stars above the threshold within the bounds of Bochum\,11 display the lowest degree of clustering behaviour of the five clusters - having a maximum of just 3-4 stars in their local neighbourhoods above random. The high/low proportion of stars within the radial boundary of Tr14/Bochum\,11 with an index value above the significance threshold, suggests these clusters are the most and least tightly clustered respectively in the region. In the absence of kinematic data however, we refrain from drawing any conclusions as to the physical origin of this trend. 

The bottom plot in Figure \,\ref{fig_carina_knownclusters} shows the distribution of stars with their index values and the positions of the 19 sub-clusters\footnote{The 20th sub-cluster, their ‘G’ cluster, is ignored in our analysis due to its large angular extension across the centre of the region} found by \citet{2014ApJ...787..107K}, as stellar overdensities using finite mixture models, overlaid. Fifteen sub-clusters are clearly identified with a significant number of members that have an index value above the defined threshold. Four sub-clusters ($F$, \textit{P}, $R$, $S$) do not contain any (or very few) stars spatially clustered above random, suggesting these sub-clusters may not be real clusterings but instead fluctuations in the dispersed population field.

We now look at the clustering tendencies of individual stars across the Carina Nebula. A total of $35.2\%$ of stars in the catalogue are clustered above random. Stars in the NW region typically have higher index values than the SE region, with $49.9\%$ and $9.1\%$ clustered above random respectively. To gauge the significance of this difference between the NW and SE regions we run a 2 sample K-S test of the index values for all stars in the NW region against those for all stars in the SE region, with a strict significance boundary of $p<0.01$, finding a value of $p<<0.001$ i.e. stars in the NW and SE regions have significantly different clustering tendencies. This result is not entirely unexpected as the NW region is heavily sub-structured, containing 3/5 of the young clusters, and 12/19 of the sub-clusters detected by \citet{2014ApJ...787..107K}  - whereas the SE region is comparatively sparsely populated and being shaped by radiative winds of the Tr14 and Tr16 clusters \citep{2008hsf2.book..138S}. Therefore the disparity of the SE and NW regions clustering tendencies is reflective of differences in the apparent star formation activity in these regions.

\begin{figure*}
\centering
   \includegraphics[width=0.45\textwidth]{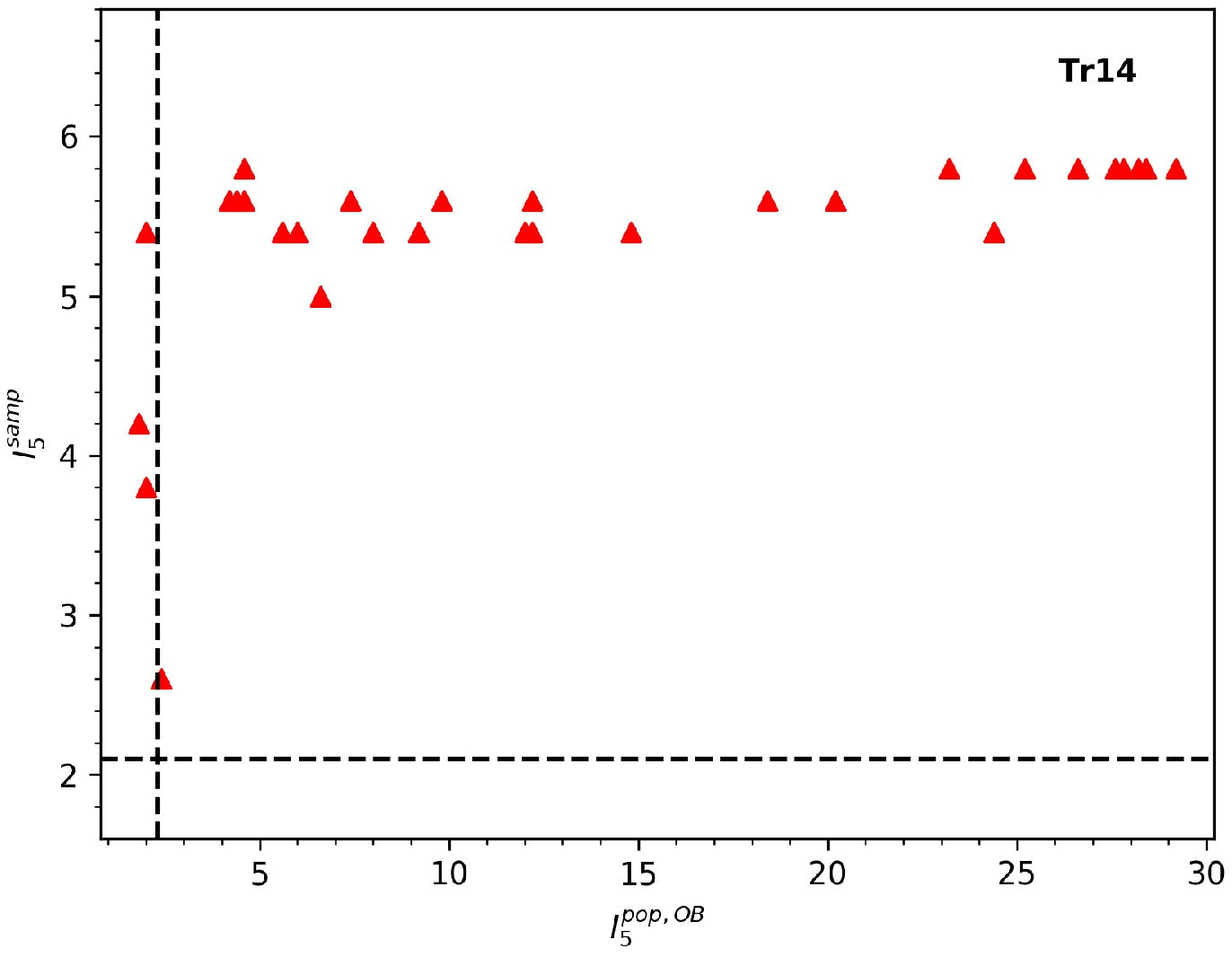} 
   \includegraphics[width=0.45\textwidth]{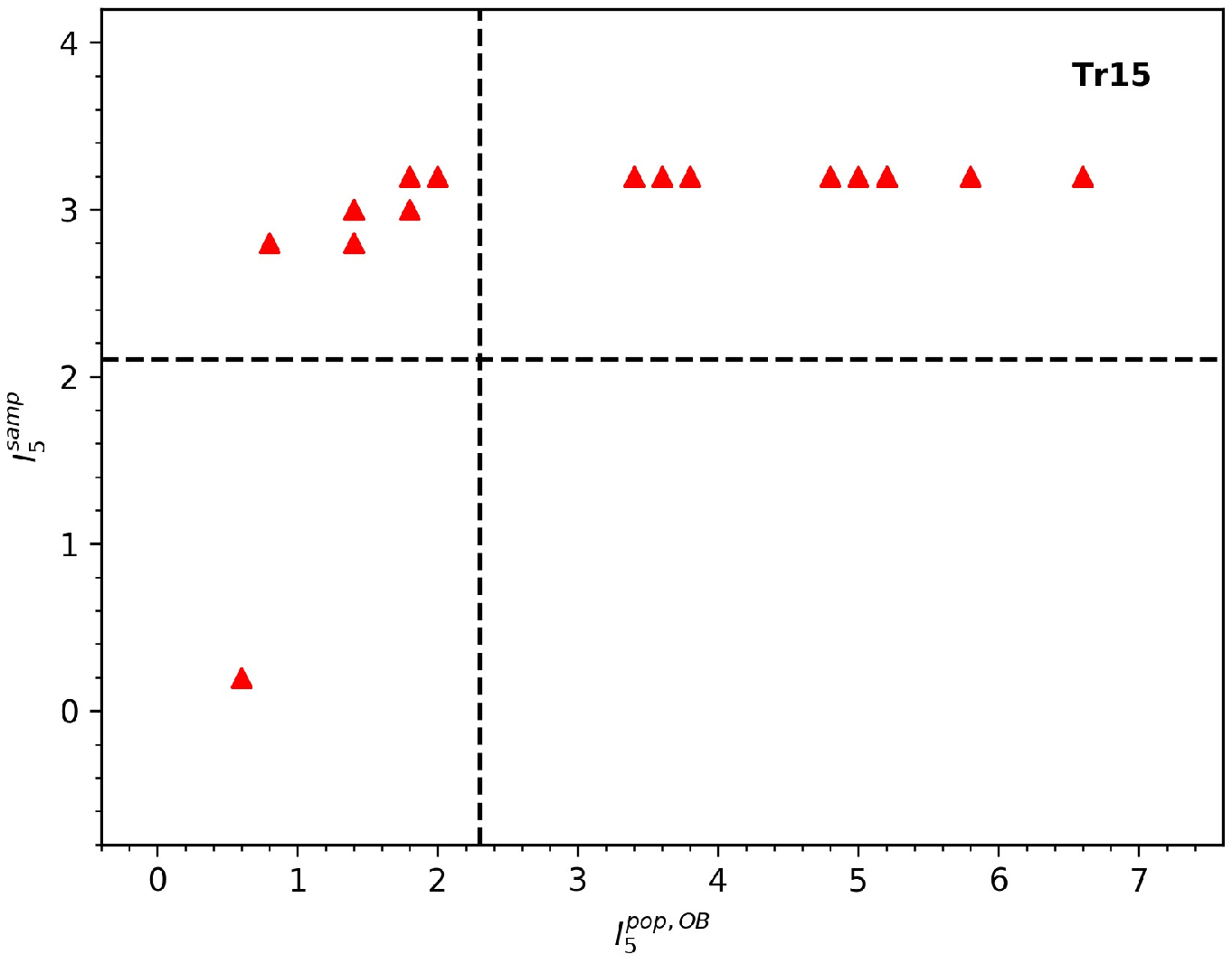} 
   \includegraphics[width=0.45\textwidth]{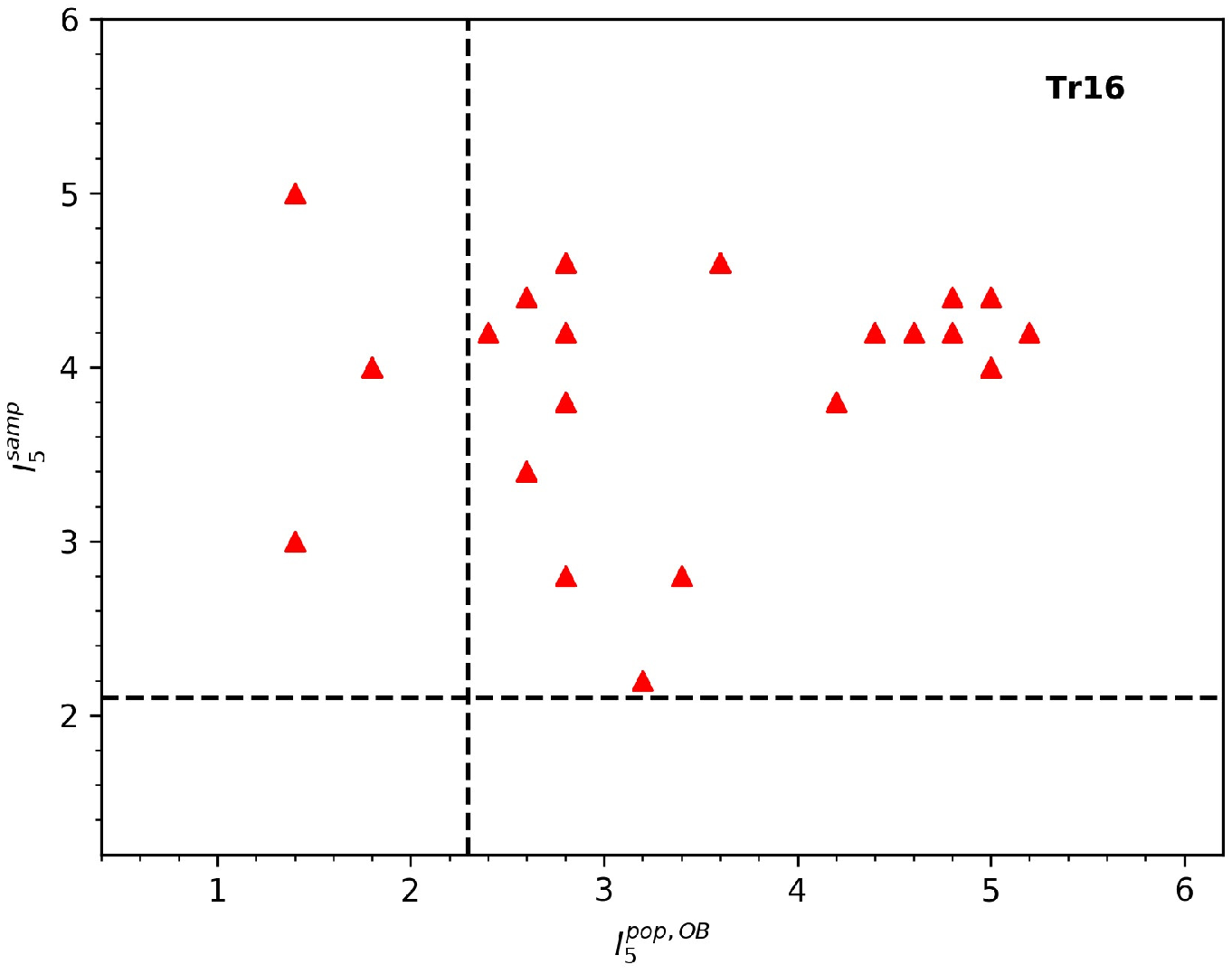} 
  \caption{Plots of index values calculated for the OB population by INDICATE when applied to the entire stellar catalogue, $I^{pop,OB}_{5}$, and the sub-sample, $I^{samp}_{5}$. Dashed black lines represent the respective significance thresholds (see text for details). Red Triangles represent OB stars within the radius of (Top Left:) Trumpler 14 (Top Right:) Trumpler 15 (Bottom:) Trumpler 16.}  \label{fig_massseg}
\end{figure*}

\subsection{Clustering Tendencies of the OB Population}\label{sect_carina_ms}

We create a sub-sample of the OB stars in the catalogue. To identify OB stars a cross-match search of the catalogue with the SIMBAD\footnote{\url{http://simbad.u-strasbg.fr/simbad/}} database was performed, finding 134 stars listed as either O or B spectral type. Thirteen of these have an ambiguous spectral type or are flagged as being higher order systems so are excluded from the sub-sample, leaving a final selection of 121 stars.

The term ‘mass segregation’ is used interchangeably in the literature to describe two quite different realisations. The first definition (hereafter \textit{Type 1}) refers to a system in which the massive stars are concentrated together at its centre; whereas the second definition (hereafter \textit{Type 2}) refers to a system in which the massive stars are in stellar concentrated regions, but are not necessarily concentrated together.

As the index quantifies the degree of association of stars, it by definition identifies (and quantitatively measures) Type 2 mass segregation as values are assigned to stars based on the degree of spatial clustering in their local neighbourhood. Figures \,\ref{fig_carina_OB}  and \,\ref{fig_carina_zoomin} show the index distribution of the OB stars with the positions of the Tr14-16, TC, Bochum\,11 and M (sub) clusters overlaid. We find the Tr14, Tr15, Tr16, TC, Bochum\,11 and M (sub) clusters have signatures of Type 2 mass segregation. A total of $57.0\%$ of OB stars are clustered above random, which is notably higher than the general populations $35.2\%$ (Sect.\,\ref{sect_struct}) i.e. cluster concentrations are more frequent around massive stars than typical for stars in this region. Massive stars in the NW region have notably different clustering tendencies to those in the SE region, with $68.1\%$ and $18.5\%$ clustered above random respectively. To gauge the significance of this difference we run a 2 sample K-S test of the index values for all sub-sample stars in the NW region against those for all sub-sample stars in the SE region, with a strict significance boundary of $p<0.01$, finding a value of $p<<0.001$, which confirms OB stars in the NW and SE regions have significantly different clustering tendencies. These results show signatures of Type 2 mass segregation are present across Carina but are primarily found in the NW region.

It is also possible to use our tool to find signals of the ‘classical’ Type 1 mass segregation. We apply INDICATE to the sub-sample of OB stars with a nearest neighbour number of $\mathit{N}=5$ using an extended control distribution (CDB) and define a significance threshold - that is the value of $I_5$ above which a star is significantly clustered above random - of three standard deviations above the mean value expected from a random distribution of the same size evaluated with  $\mathit{N}=5$ and CDB, such that $I_{sig}=\bar{I}^{random}_5+3\sigma = 2.1$.  As the index is a quantitative measure of the degree of clustering of OB stars with other OB stars, it is a local measure of Type 1 mass segregation. We find the massive population is notably more self-clustered than is typical amongst the general stellar population with a total of $64.5\%$ of stars in the sub-sample clustered above random ($\sim$ a factor of two larger than for the general population). The Tr14, Tr15 and Tr16 clusters have signatures of Type 1 mass segregation with a significant number of OB members more clustered than expected for a random distribution, and mean index values of OB stars within their cluster radii of $I^{\,samp}_{5}= 5.4$, $3.0$ and 3.9 respectively. Neither the TC or Bochum\,11 clusters have signatures of mass segregation, with mean index values of $0.2$ and $0.8$ respectively. Massive stars in the NW region have completely different clustering tendencies than the SE region, with $83.0\%$ and $0.0\%$ clustered above random respectively. To gauge the significance of this difference we run a 2 sample K-S test of the index values for all sub-sample stars in the NW region against those for all sub-sample stars in the SE region, with a strict significance boundary of $p<0.01$, finding a value of $p<<0.001$, which confirms OB stars in the NW and SE regions have significantly different clustering tendencies. These results clearly show that signatures of Type 1 mass segregation are present in the NW region but not in the SE region - massive stars here are not spatially concentrated together above random. 

Finally, we look for correlations in the clustering behaviour of OB stars - is there a relation between the stellar concentrations around massive stars and the self-concentration of the OB population in the Carina region? Figure\,\ref{fig_massseg} shows a comparison of the index values derived for the OB population of Tr14, Tr15 and T16 from the application of INDICATE to (1) the entire stellar catalogue (Sect.\,\ref{sect_struct}) and (2) the OB sub-sample (Sect.\,\ref{sect_carina_ms}). In both Tr14 and Tr15 there is a clear trend between the concentration of OB stars and the concentration of (lower mass) stars around OB stars: while there is a maximum degree of association an OB star can have w.r.t. other OB stars, stellar concentrations around an OB star may continue to increase. We find that Tr16 does not follow this trend, which is consistent with what is known about the structure of the Trumpler clusters. Unlike the Tr14 and Tr15 clusters, Tr16 does not have a strong central concentration but instead is irregularly shaped and heavily sub-structured with multiple sub-clusters (\citealt{2007A&A...476..199A}, \citealt{2011ApJS..194...11W}, \citealt{2011ApJS..194...12W}). Thus the index values of Tr16 reflect that the OB stars are not clustered together in a single concentration with a (near) constant degree of clustering, but are instead scattered across a region with local concentrations of stars and a variable degree of association.

%######################################################################################################
\section{Conclusions}\label{sec_conclude}

We have developed a powerful novel statistical clustering tool called “INDICATE” (INdex to Define Inherent Clustering And TEndencies) to study the intensity, correlation and spatial distribution of point processes in discrete astronomical datasets. The tool assesses the clustering tendency of each object in a dataset and assigns it an index $I_{j, \mathit{ N}}$ (Eq.\,\ref{eq_I}), using a nearest neighbour approach by comparing the spatial distribution of objects in its local neighbourhood with that expected in an evenly spaced uniform (i.e. definitively non-clustered) distribution. INDICATE requires no a priori knowledge of, and makes no assumptions about, the shape of a distribution, presence/number of clusters and/or sub-structure of a dataset. 

For any application of INDICATE there are three variable parameters: (1) size and (2) number density of the distribution it is being applied to; and (3) the N$^{th}$ nearest neighbour number used by the tool. We calibrated our tool against random distributions to define statistically significant values of $I_{j, \mathit{ N}}$ (see appendices) finding: 

\begin{enumerate}

\item There is a logarithmic relationship between the maximum $I_{j, \mathit{ N}}$ value for a random distribution and sample size (Eq.\,\ref{eq_Imax}). \newline

\item $I_{j, \mathit{ N}}$ is independent of a distributions number density. \newline

\item There is a relationship between the typical modal $I_{j, \mathit{ N}}$ value for a random distribution and the chosen N$^{th}$ nearest neighbour number (Eq.\,\ref{eq_Imo}). \newline

\item The size of the control distribution is essentially arbitrary, as the modal difference between $I_{j, \mathit{ N}}$ values calculated for a random distribution using the standard size (Sect.\,\ref{sect_method_descrip}) and an expanded size (Appendix\,\ref{sec_edge}) is inversely proportional the the N$^{th}$ nearest neighbour number (Eq.\,\ref{eq_NN}). However, care should be taken when including/excluding points which are on the boundary of a chosen significance threshold value of $I_{j, \mathit{ N}}$ based solely on their index values during an analysis, particularly for indices derived for small sample sizes using large nearest neighbour numbers. \newline

\item Uniformly distributed (interloping) field stars in observational datasets typically do not significantly affect the index values of true cluster members. The error on the
index value derived for true cluster members is given by Eqs.\,\ref{eq_Igrad_range}, \ref{eq_Ierr_F1} and \ref{eq_Ierr_F2}.\newline

\item If interloping field stars are distributed in a gradient, the index derived for true cluster members is independent of gradient shape for small nearest neighbour numbers ($\mathit{ N}=3$). However, as field stars are also assigned index values, care must be taken when drawing conclusions on the physical origins of the clustering tendencies of stars in the dataset.\newline

\end{enumerate}

One of the primary strengths of our tool is its versatility and flexibility to be applied to a user-defined analysis. In this paper we demonstrated one potential application of the tool - to look for signals of mass segregation and trace variations in degree of stellar association in star forming regions/clusters. 

Arguably the three most popular established methods to identify mass segregation are: (i) Radial Mass Functions (e.g. \citealt{1988MNRAS.234..831S}), (ii) the $\Lambda_\mathrm{MSR}$ parameter \citep{2009MNRAS.395.1449A}, and (iii) the Local Density Ratio (\citealt{2011MNRAS.416..541M}; \citealt{2011MNRAS.417.2300K}; \citealt{2014MNRAS.445.4037P}). Each have their respective strengths and weaknesses (see \citealt{2015MNRAS.449.3381P} for a discussion), but primarily the decision of which method one employs is based upon what type of mass segregation one is searching for. In the literature, ‘mass segregation’ is used interchangeably to describe two quite different realisations: (1) the concentration of massive stars together at a system's centre and (2) (lower mass) stellar concentrations around the massive stars in a system but which are not necessarily concentrated together. In our case, we are interested in better understanding the role of local and global environmental conditions in massive star formation. Our aim therefore was to measure the degree of association (or lack thereof) of each high mass star with the general stellar population and with each other, in young  ($<$\,5\,Myr) regions, i.e. look for signatures of both types of mass segregation. For our purpose, it is ideal to employ a single method to search for and quantify signatures of both types in a given region, so that they can be directly compared and a quantitative analysis of the impact of local environment formation conditions on spatial structures undertaken. This is possible using INDICATE, as demonstrated in  Sect.\ref{sect_carina_ms}.

Additional strengths of INDICATE are: \newline

\begin{itemize}

\item Our tool does not require a priori knowledge of the centre, and works independently of the shape, of the distribution. \newline 

\item The index has been calibrated against random distributions, so statistically significant values are easily identified. \newline

\item As $I_{j,\mathit{ N}}$ is a measure of spatial association (not density), the clustering behaviour (index values) of massive stars in two or more regions can be directly compared, regardless of differences in their distances, average angular separation of sources and/or field sizes. \newline 

\item It can provide both a global \textit{and} local measure of Type 1 mass segregation. By definition $I_{j,\mathit{ N}}$ is a local measure, and a global measure can be obtained for the subset by e.g. calculating the mean index value of the massive stars and comparing it to that expected by a random distribution (Sect.\ref{sect_carina_ms}). \newline 

\item Conclusions on Type 1 mass segregation in a system are not based on the larger spatial distribution of other stars in the system as a whole. Index values for high mass stars are derived through comparison to a control distribution, not internally with other sub-samples of the system (low mass stars), and significant values are determined through comparison to those expected in a random distribution. Therefore high mass stars index values are independent of the completeness of the resolved low mass population census. \newline

\item As a local measure, INDICATE is robust against outliers as they (a) will not influence the index values of the other members in a subset, (b) are easily identifiable by their comparative low index values and as such (c) in an global analysis of a system to find signatures of Type 1 mass segregation will have a statistically negligible effect on the overall conclusions drawn for the subset. \newline

\end{itemize}

We applied our tool to the stellar catalogue of the Carina Nebula (NGC\,3372) by \citet{2014ApJ...787..107K}, a region chosen because of its known high mass stellar content ($>$130 OB stars) and extensive sub-structure:

\begin{enumerate}

\item We recover known stellar structure in the region, including the Tr14-16, Treasure Chest and Bochum\,11 clusters. \newline

\item We find members of the 4/19 sub-clusters identified by \citet{2014ApJ...787..107K} as stellar overdensities are more clustered than typical for the extended distribution of stars in the Carina region, but contain no, or very few, stars with a degree of association above random. This suggests these sub-clusters may be fluctuations in the dispersed population field rather than real clusters. \newline

\item Stars in the NW and SE regions have significantly different clustering tendencies. The NW region is known to be heavily sub-structured, whereas the SE is more sparsely populated and being shaped by radiative winds of the Tr14 and Tr16 clusters \citep{2008hsf2.book..138S}. Therefore this result is reflective of differences in the apparent star formation activity in these regions. Further study is required to ascertain the physical origin of that difference. \newline

\item The different clustering properties between the NW and SE regions are also seen for OB stars and are even more pronounced. \newline

\item There are no signatures of classical (Type 1) mass segregation present in the SE region - massive stars here are not concentrated together above random. \newline  

\item Stellar concentrations are more frequent around massive stars than typical for the general population, particularly in the young Tr14 cluster. \newline 

\item For Tr14 and Tr15 we find a relation between the concentration of OB stars and the concentration of (lower mass) stars around OB stars. This relation is notably absent from Tr16.  Unlike the Tr14 and Tr15 clusters, Tr16 does not have a strong central concentration but instead is irregularly shaped and heavily sub-structured with multiple sub-clusters (\citealt{2007A&A...476..199A}, \citealt{2011ApJS..194...11W}, \citealt{2011ApJS..194...12W}). Therefore this result reflects the known structure of the clusters: Tr14 and Tr15 are centrally concentrated, whereas in Tr16 the OB stars are not clustered together in a single concentration with a (near) constant degree of clustering, but are instead scattered across a region with local concentrations of stars and a variable degree of association.  \newline

\end{enumerate}

%######################################################################################################
\begin{acknowledgements}

The Star Form Mapper project has received funding from the European Union’s Horizon 2020 research and innovation programme under grant agreement No 687528. We would like to thank Simon Goodwin and Lee Mundy for suggestions for the improvement of our tool and paper respectively.

\end{acknowledgements}

% for the bibliography, at the end
\bibliographystyle{aa} % style aa.bst
\bibliography{refs} % your references Yourfile.bib

\begin{appendix}

\section{Calibration of the Index}\label{sec_cali}

We conduct a series of baseline tests to aid interpretation and identification of significant index values. These tests (a) define the threshold at which an index value becomes significant i.e. the value above which it can reasonably be assumed point $j$ was not drawn from a random distribution and (b) quantify the impact of dataset parameters and the choice of N$^{th}$ nearest neighbour number on the distribution and range of index values INDICATE generates.

\begin{enumerate}[($i$)]
\item \textit{Sample Size} \newline 

We generate random samples of size, $S$, in the range $50\,\leq\,\mathrm{S}<\,100,000$. For each sample size 100 realisations are created with a constant number density of $n_\mathrm{obs}=1$ object per unit area and INDICATE is implemented with a N$^{th}$ nearest neighbour number of $\mathit{N}=5$. We keep the number density and N$^{th}$ nearest neighbour number for every sample constant to ensure any identified trends or patterns in samples’ index values can be attributed to sample size alone.  \newline 

There is no dependence between the index and the size of a sample, with typical modal and mean values of Mo$[I_5] =0.8$ and $\bar{I_5} =1.0$ for random distributions under the above stated conditions. However, we find there is a logarithmic relationship between the upper range limit\footnote{derived as the maximum value over all realisations} of $I_{j,\mathit{ N}}$ and sample size for random distributions, i.e. 

			\begin{equation}
			\\ 0.0 \leq I_{j,\mathit{ N}} \leq I_\mathrm{max}
			\end{equation}

where

			\begin{equation}\label{eq_Imax}
			\\ I_\mathrm{max} = C_\mathrm{1}+C_\mathrm{2}\times\log_{10}\mathrm{S}
			\end{equation}

and $C_\mathrm{1}, C_\mathrm{2}$ are constants which are dependant on the N$^{th}$ nearest neighbour number (see Table\,\ref{table_Imax}). Equation\,\ref{eq_Imax} defines as a function of sample size the threshold value above which we can definitively assume a point does not have a spatially random distribution. \newline

\item \textit{Field Density} \newline 

We generate random samples of number density, $n_\mathrm{obs}$, in the range $10^{-6} \leq n_\mathrm{obs} \leq 10^{6}$, in increments of an order of magnitude. For each value of number density 100 realisations are created, with a constant sample size of $10,000$ and INDICATE is implemented with a N$^{th}$ nearest neighbour number of $\mathit{N}=5$. We find there is no dependence of $I_{j,\mathit{ N}}$ on field density. \newline  

\item \textit{N$^{th}$ Nearest Neighbour Number, $\mathit{N}$} \newline

We generate a 100 realisations of random samples of size $S=10,000$ and number density $n_\mathrm{obs} =1$. For each sample INDICATE is implemented with a N$^{th}$ nearest neighbour number of $\mathit{N}=3$, $5$, $7$ and $9$. There is a relationship between the upper range limit of $I_{j,\mathit{ N}}$, sample size and N$^{th}$ nearest neighbour number (Eq\,\ref{eq_Imax}, Table\,\ref{table_Imax}). The typical modal index value, $\text{Mo}[I_{j,\mathit{ N}}]$, of randomly distributed samples vary as a function of $\mathit{N}$:

			\begin{equation}\label{eq_Imo}
			\\ \text{Mo}\left[I_{j,\mathit{ N}}\right] \equiv \frac{\mathit{N}-1}{\mathit{N}}
			\end{equation}

The typical mean index values of randomly distributed samples are  $0.9\le\,\bar{I}_{\mathit{N}}\le\,1.0$. 

\end{enumerate}

\begin{table} 
\caption{Constants of Eq.\,\ref{eq_Imax} for a N$^{th}$ nearest neighbour number of $\mathit{N} =$ 3, 5, 7 and 9 with their respective fit correlation coefficient ($\mathrm{R}$) and standard error (\textit{SE}). }              % title of Table
\label{table_Imax}      % is used to refer this table in the text
\centering                                      % used for centering table
\begin{tabular}{c c c c c}          % centered columns (5 columns)
\hline\hline                        % inserts double horizontal lines
$\mathit{N}$ & $C_\mathrm{1}$ & $C_\mathrm{2}$ & $\mathrm{R}$ & \textit{SE} \\    % table heading
\hline                                   % inserts single horizontal line
    3 & 2.508 & 0.489 & 0.831 & 0.047\\      % inserting body of the table
    5 & 2.291 & 0.361 & 0.851 & 0.032 \\
    7 & 2.244 & 0.206 & 0.796 & 0.023 \\
    9 & 2.093 & 0.197 & 0.777 & 0.023 \\
\hline                                             %inserts single line
\end{tabular}
\end{table}

\section{Investigation of Edge Effects}\label{sec_edge}

\begin{figure}
\centering
   \includegraphics[width=0.5\textwidth]{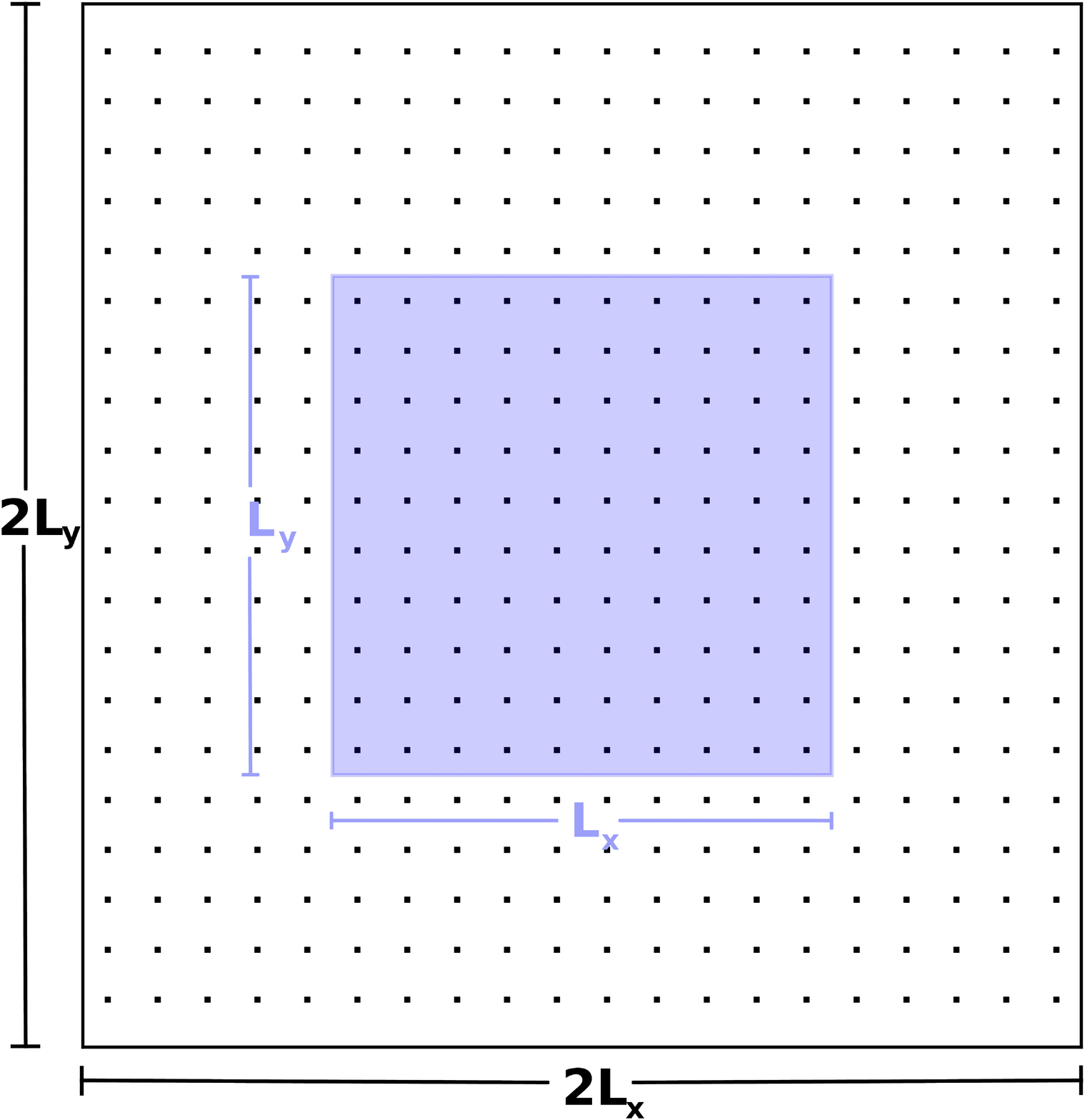}
  \caption{Dimensions of control distributions CDA (blue shaded) and CDB (all visible) as described in Appendix\,\ref{sec_edge}, where $L_{x}$ and $L_{y}$ are the length of a test sample's x and y axis respectively; and the black dots are the points of the control distributions.}  \label{fig_expandcontrol}
\end{figure}

Section 2.1.2 described how the control distribution used by INDICATE is generated. Here we investigate whether the proximity of a point in a dataset to its delimited boundaries and/or the total length of each axis of the control distribution influences a sample's index values. We repeat the calibration tests (Appendix \,\ref{sec_cali}) using two different types of control distribution:

\begin{enumerate}

\item \textit{Control Distribution A} (CDA) - occupies the same bounded parameter space and has the same number density (Eq.\,\ref{eq_ncon}) as the test sample; \newline

\item \textit{Control Distribution B} (CDB) - occupies the same, and is extended beyond the, bounded parameter space of the test sample; such that area of the control distribution is a factor of four times larger than the test sample (see Figure\,\ref{fig_expandcontrol}). Increasing the area of the control distribution by a factor of four ensures that the $r_j$ of edge points in the test sample (Eq.\,\ref{eq_rj}) is not calculated using edge points of the control distribution (which in principle could subsequently increase $\bar{r}$, and decrease $I_{j,\mathit{ N}}$). It has the same number density as the test sample (Eq.\,\ref{eq_ncon}).\footnote{For samples with non-rectangular delimited areas this distribution should always be used.}

\end{enumerate}

We define an `edge point' as any point in the sample dataset whose (x,y) position is less than that of the second smallest x and/or y positions and/or greater than the second largest x and/or y positions of points in CDA (i.e. where the measured nearest neighbour distance of the sample point to the control distribution points would be affected due to lack of control points in any given direction in the control distribution for point $j$). 

For $\mathit{N}>5$ the modal index value of edge points, $\text{Mo}[I_{j,\mathit{ N}}^{\mathrm{\,E}}]$, deviates from that of the sample as a whole using CDA (Eq.\,\ref{eq_Imo}), such that

			\begin{equation}\label{eq_Imo_nn7}
			\\ \text{Mo}\left[ \,\,I_{j, 7}^{\mathrm{\,E}}\,\, \right] \equiv \frac{\mathit{N}-2}{\mathit{N}}\,\,\,\,\,\,\text{for}\,\,\,\,\,\,\,\,\,\,\mathit{N} = 7\,\, 
			\end{equation}

			\begin{equation}\label{eq_Imo_nn9}
			\\ \text{Mo}\left[ \,\,I_{j, 9}^{\mathrm{\,E}}\,\, \right] \equiv \frac{\mathit{N}-3}{\mathit{N}}\,\,\,\,\,\,\text{for}\,\,\,\,\,\,\,\, \,\,\mathit{N} = 9\,\, 
			\end{equation}

For a proportion of all (edge and non-edge) points in the test samples’ there is a statistically small discrepancy between the index values calculated using CDA and CDB. The modal difference between the two sets of indices is inversely proportional to $\mathit{N}$ i.e. 

	\begin{equation}\label{eq_NN}
	\\ \text{Mo}\left[ \,\,\Delta{I_{j,\mathit{ N}}}\,\, \right]\,\,\equiv\,\,\text{Mo}\left[ \,\,\,I^\mathrm{\,CDA}_{j, \mathit{N}} - I^\mathrm{\,CDB}_{j, \mathit{N}}\,\, \right]\,\,\equiv\,\,\frac{1}{\mathit{N}}  \,\,\text{for}\,\, \,\,\mathit{N} \ge 3\,\, 
	\end{equation}

where $I^\mathrm{\,CDA}_{j, \mathit{N}}$ and $I^\mathrm{\,CDB}_{j, \mathit{N}}$ are the index values calculated for each sample point $j$ using CDA and CDB respectively. For any given point $j$ if 

	\begin{equation}
	\\ \Delta{I_{j,\mathit{ N}}}\,\,>\,\,0 \,\,\,\,\,\,\leftrightarrow\,\,\,\,\,\,\,\,I^\mathrm{\,\,CDA}_{j, \mathit{N}} > I^\mathrm{\,\,CDB}_{j, \mathit{N}}
	\end{equation}

The proportion of all points with $\Delta{I_{j,\mathit{ N}}} > 0$ increases with decreasing sample size and increasing nearest neighbour number, reaching $\sim 90\%$ for sample size of $\mathrm{S}=50$ using $\mathit{N}=9$; it is independent of field density. The number of edge points with $\Delta{I_{j,\mathit{ N}}}> 0$ is proportionally lower than non-edge points i.e. expanding the control distribution has less of an effect on edge points than non-edge points. This is because the $r_j$ measured for edge points in CDB is (slightly) smaller than in CDA (as it is no longer artificially increased due a lack of control points in any given direction), which subsequently causes a small decrease in $\bar{r}$ (Eq.\,\ref{eq_rbar}). In both control distributions a radius of $\bar{r}$ from an edge point can partially encompass an area outside the bounds of the dataset (where there can be no neighbouring points), but for non-edge points a radius of $\bar{r}$ always encompasses an area within the bounds of the dataset (neighbouring points can be present in any given direction within $\bar{r}$). Thus a small decrease in $\bar{r}$ is more likely to exclude a nearest neighbour (decrease $N_\mathrm{\bar{r}}$, and subsequently $I_{j,\mathit{ N}}$ - Eq.\,\ref{eq_I}) for a non-edge point than an edge point. 

To conclude, as the typical $\Delta{I_{j,\mathit{ N}}}$ for any given point between the two control distributions is very small, choice of control distribution type (CDA or CDB) is essentially arbitrary, but care should be taken when including/excluding points which are on the boundary of a chosen significance threshold value of $I_{\mathit{N}}$ during an analysis - particularly indices derived for small sample sizes using large nearest neighbour numbers.

%~~~~~~~~~~~~~~~~~~~~~~~~~~~~~~~~~~~~~~~~~~~~~~~~~~~~~~~~~~~~~~~~~~~~

\section{Investigation of Field Effects}\label{sec_inter}

\begin{figure*}
\centering
   \includegraphics[width=6cm,height=5cm]{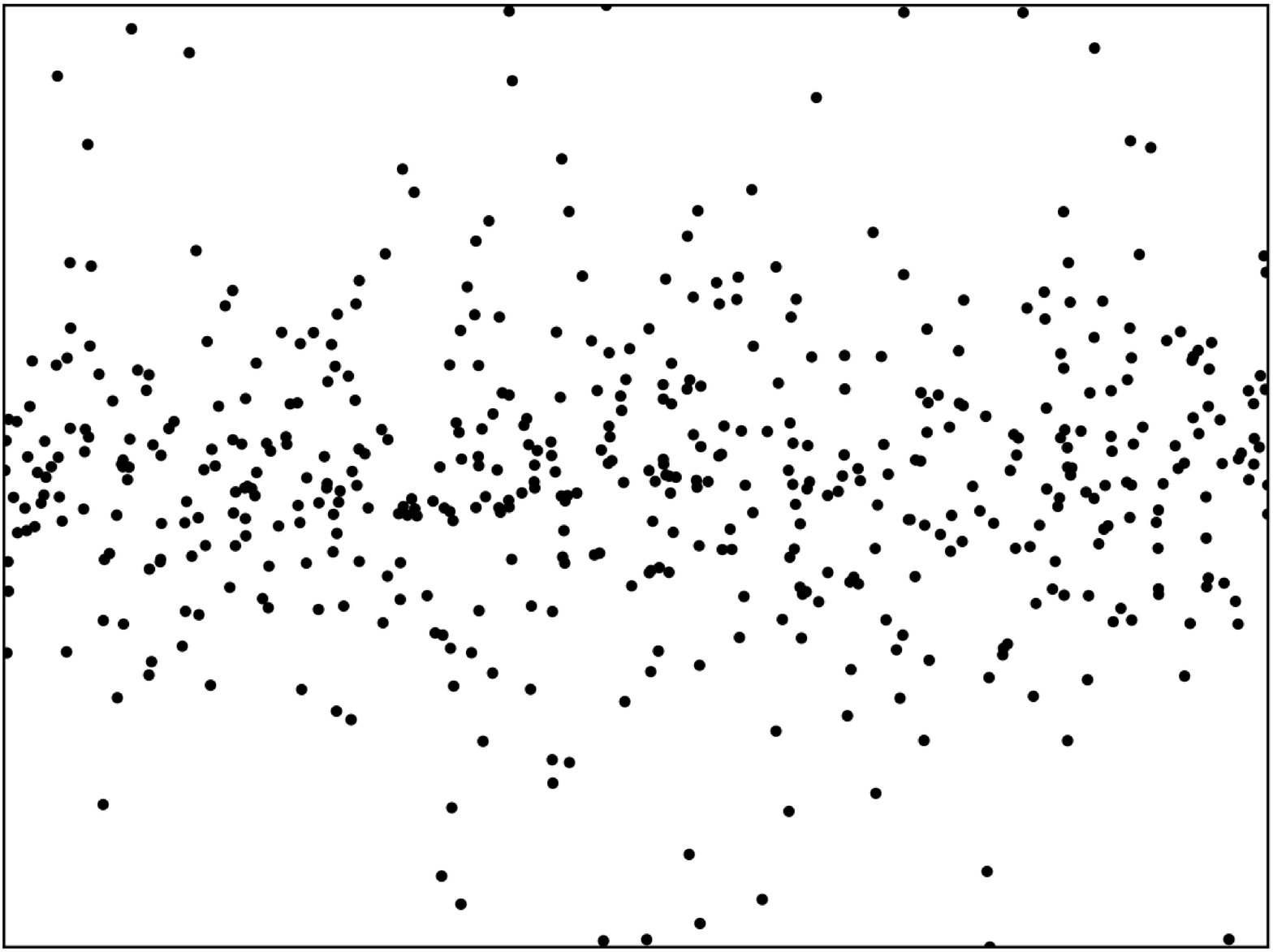} \hfill
   \includegraphics[width=6cm,height=5cm]{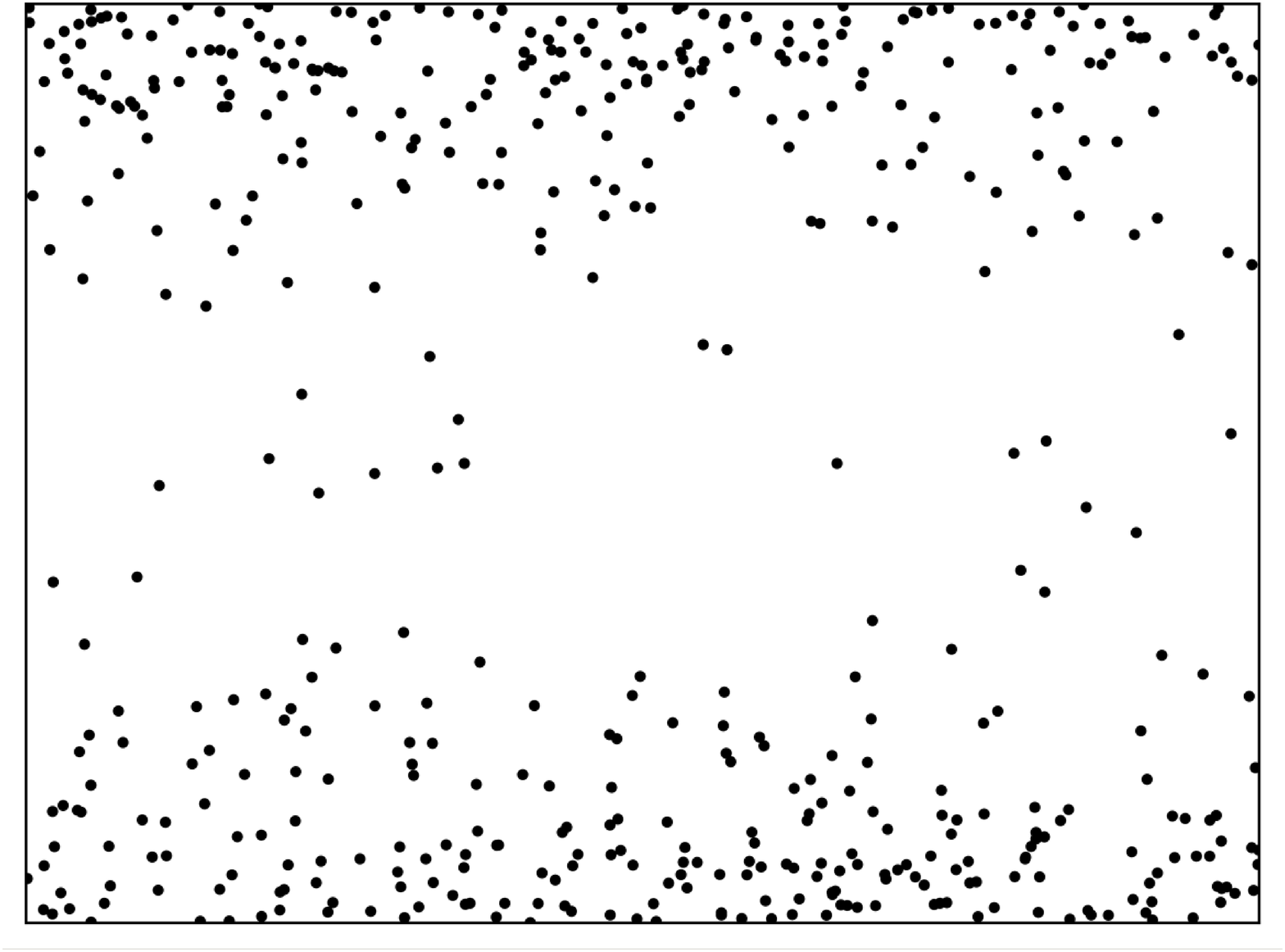} \hfill
   \includegraphics[width=6cm,height=5cm]{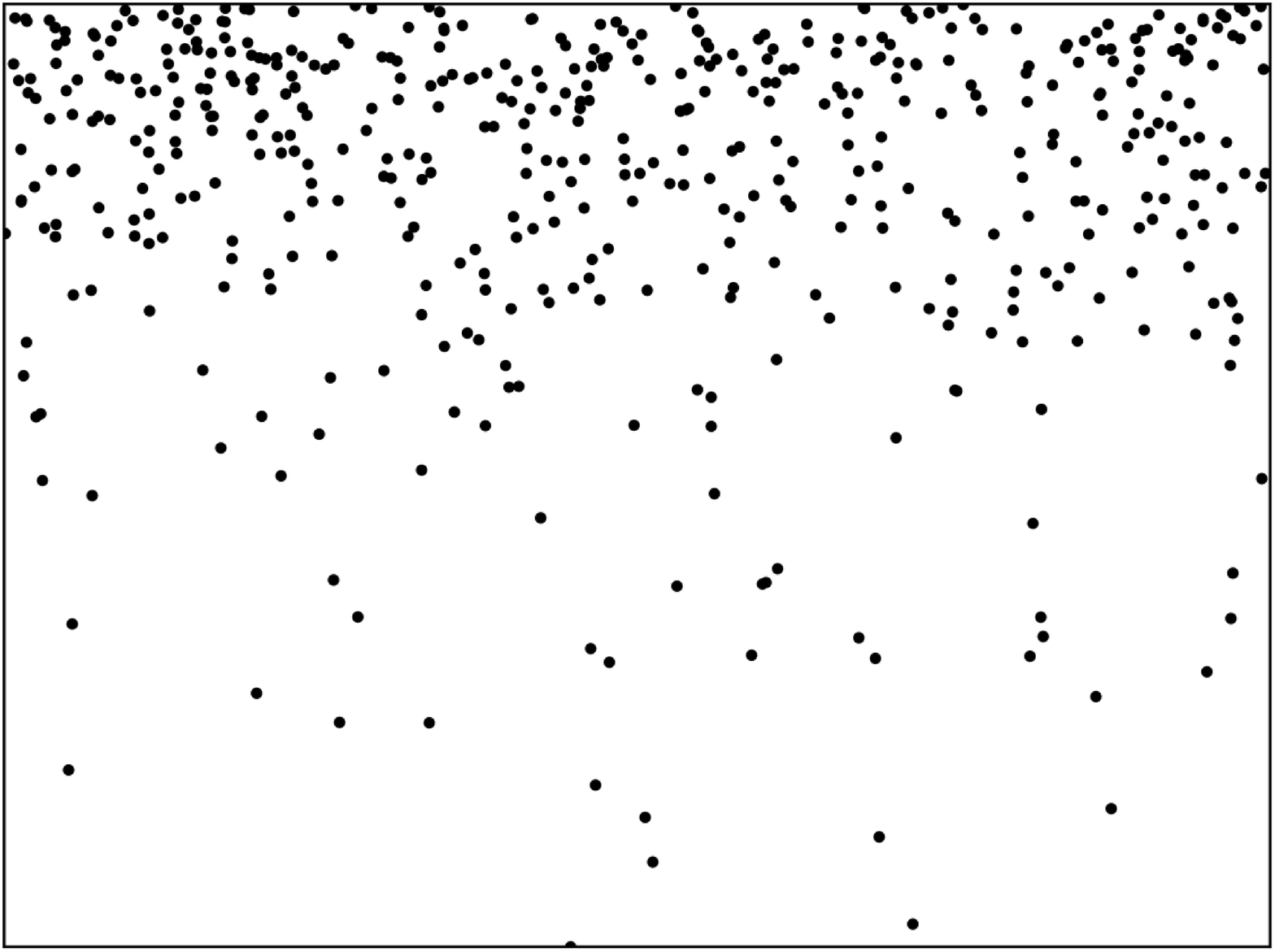} \hfill
  \caption{Plots show a realisation of the three gradient field population shapes tested in Appendix\,\ref{sec_inter}.}  \label{fig_gradient}
\end{figure*}

To ascertain the influence of interlopers on the index values of true cluster members we conduct additional calibration tests. A dataset consisting of a Gaussian cluster with 500 members is generated and the index value of each member determined using the steps outlined in Section\,\ref{sect_method_descrip}. 

In our first test, field stars are introduced to the dataset with incrementally increasing frequency, such that the number of interloping field stars at any given time is equal to a fraction, $F$, of total cluster members in the range $0.01\le\,F\le\,1.0$. The positions of the field stars are randomly drawn from a uniform distribution. For each fraction of field stars 100 realisations are made, and for each realisation the difference, $\Delta I_{j,\mathit{ N}}$ , between the index values derived for cluster members in the dataset that does not contain field stars and the current level of field star contamination is measured for a N$^{th}$ nearest neighbour number of $\mathit{ N}$ = 3, 5, 7 and 9. As we are simulating an observational dataset for which cluster membership is uncertain, $N_\mathrm{tot}=S=500+(F\times500)$.

We find the modal difference for all combinations of $F$ and $\mathit{ N}$ is Mo$[\Delta I_{j,\mathit{ N}}]= 0$, i.e. typically the index values of true cluster members are unaffected by the presence of interloping field stars. The proportion of cluster members with $\Delta I_{j,\mathit{ N}} \ne 0$ increases with increasing $F$ and $\mathit{ N}$, reaching a maximum of $\sim 95\%$ for $F=1.0$ and $\mathit{N}=9$. In observationally obtained datasets the error on the index value derived for true cluster members is therefore

	\begin{equation}\label{eq_Igrad_range}
	\\ I_{j, \mathit{ N}}+F2\,\le\,I_{j, \mathit{ N}}\,\le\,\,I_{j, \mathit{ N}}+F1
	\end{equation}
	
	where 

	\begin{equation}\label{eq_Ierr_F1}
	\\ F1=\max\left[\Delta I_{j,\mathit{ N}}\right]= C_{3}+C_{4}\,\times\,\log\left(F\right)
	\end{equation}

	\begin{equation}\label{eq_Ierr_F2}
	\\ F2= \min\left[\Delta I_{j,\mathit{ N}}\right]= C_{5}\,\times\, \exp\left(C_{6}\,\times\,F\right) + C_{7}
	\end{equation}

and $C_{3-7}$ are constants dependant on the N$^{th}$ nearest neighbour number (see Table \ref{table_Ierr}). %, .

In our second test $F=1.0$ field stars are distributed in three large scale gradient patterns (Figure\,\ref{fig_gradient}) which are randomly generated in the same parameter space as the Gaussian cluster. For each gradient 100 realisations are made, and for each realisation the difference, $\Delta I_{j,\mathit{ N}}$ , between the index values derived for cluster members in the dataset that does not contain field stars and the current level of field star contamination is measured for a N$^{th}$ nearest neighbour number of $\mathit{ N}$ = 3, 5, 7 and 9. As we are simulating an observational dataset for which cluster membership is uncertain, $N_\mathrm{tot}=S=500+(F\times500)=1000$.

We find the modal difference for all gradients with $\mathit{N}=3$ is Mo$[ \Delta I_{j,\mathit{ N}}]= 0$, i.e. for small values of $\mathit{N}$ the index derived for cluster members is independent of gradient shape. This is expected as the index is a local measure, and the value of $\mathit{N}$ essentially defines its resolution (the smaller $\mathit{N}$, the higher the resolution). Thus index values are more susceptible to the effects of variation in the degree of field star association within the gradient when larger values of $\mathit{N}$ are employed.

As noted previously, INDICATE is distance independent for a fully
  resolved dataset.  However, in practice, clearly INDICATE cannot detect
  unresolved binaries and higher order systems in datasets nor {\em a priori} know any difference between a member of a grouping and a
  fore- or background field star.  Even with best efforts, not all field stars
  will be removed from observationally obtained datasets before analysis, so
  consideration must be given before drawing conclusions about the clustering
  tendencies of region stars. In particular, when a pronounced large scale 2D spatial distribution gradient of the field population is present, and cluster membership is uncertain, caution must be taken when drawing conclusions about the physical origins of the clustering tendencies of stars – as field stars within the denser regions of the gradient naturally will have a higher degree of association and thus index. Similar care must be taken when interpreting index values for 2D datasets in which a smaller angular resolution cluster is superimposed onto a larger angular resolution cluster, or that contains two clusters at significantly different distances. Simulations and bootstrapping techniques can be used to test the magnitude of such effects on individual datasets.

\begin{table}
\caption{Constants of Eqs.\,\ref{eq_Ierr_F1},\,\ref{eq_Ierr_F2} for a N$^{th}$ nearest neighbour number of
$\mathit{N}$ = 3, 5, 7 and 9.  \label{table_Ierr}}              % title of Table
\label{table_example}      % is used to refer this table in the text
\centering                                      % used for centering table
\begin{tabular}{c c c c c c}          % centered columns (6 columns)
\hline\hline                        % inserts double horizontal lines                                   % inserts single horizontal line
$N$ &  $C_{3}$ & $C_{4}$& $C_{5}$ & $C_{6}$ & $C_{7}$ \\ 
\hline 
3 & 2.549 &  0.510 & 4.725 & 0.830 & -6.077 \\
5 & 2.033 & 0.438  & 3.394 & 2.133 & -3.844 \\
7 & 1.617 & 0.320 & 3.715  & 1.467 & -4.140 \\
9 & 1.421 & 0.306 & 3.156  & 1.867 & -3.559\\
\end{tabular}
\end{table}

\end{appendix}

\end{document}